\documentclass[fleqn,usenatbib]{mnras}

\usepackage[T1]{fontenc}

\DeclareRobustCommand{\VAN}[3]{#2}
\let\VANthebibliography\thebibliography
\def\thebibliography{\DeclareRobustCommand{\VAN}[3]{##3}\VANthebibliography}

\usepackage{graphicx}	
\usepackage{xcolor}
\usepackage{amsmath}	
\usepackage{amssymb}	
\usepackage{subcaption}

\usepackage{booktabs} 
\usepackage{multirow,threeparttable,longtable}  

\usepackage{newtxtext,newtxmath}

\newcommand{\CORRS}[1]{{#1}}

\def\xmm{XMM-\textit{Newton}}

\title[Low-frequency radio observations of Abell 3266]{The merging galaxy cluster Abell~3266 at low radio frequencies}

\author[S.~W.~Duchesne et al.]{
S.~W.~Duchesne,$^{1,2}$\thanks{E-mail: stefan.duchesne.astro@gmail.com}
M.~Johnston-Hollitt,$^{3}$
C.~J.~Riseley,$^{4,5,1}$
I.~Bartalucci$^{6}$
and S.~R.~Keel$^{2}$
\\
$^{1}$CSIRO Space \& Astronomy, PO Box 1130, Bentley, WA 6102, Australia\\
$^{2}$International Centre for Radio Astronomy Research (ICRAR), Curtin University, Bentley, WA 6102, Australia\\
$^{3}$Curtin Institute for Computation, Curtin University, GPO Box U1987, Perth, WA 6845, Australia\\
$^{4}$Dipartimento di Fisica e Astronomia, Universit\`a degli Studi di Bologna, via P. Gobetti 93/2, 40129 Bologna, Italy\\ 
$^{5}$INAF -- Istituto di Radioastronomia, via P. Gobetti 101, 40129 Bologna, Italy\\ 
$^{6}$INAF -- Istituto di Astrofisica Spaziale e Fisica Cosmica di Milano, Via A. Corti 12, 20133 Milano, Italy\\
}

\date{Accepted XXX. Received YYY; in original form ZZZ}

\pubyear{2021}

\begin{document}
\label{firstpage}
\pagerange{\pageref{firstpage}--\pageref{lastpage}}
\maketitle

\begin{abstract}
We present new low-frequency ($\nu = 88$--216~MHz) observations of the complex merging galaxy cluster Abell~3266. These new observations are taken with the Murchison Widefield Array (MWA) in its Phase II `extended', long-baseline configuration, offering the highest-resolution low-frequency view of the cluster to date. We report on the detection of four steep spectrum ($\alpha \lesssim -1$ for $S_\nu \propto \nu^\alpha$) extended radio sources within the cluster. We confirm the detection of a $\sim 570$~kpc radio relic to the south of the cluster, and a possible bridge of emission connecting the relic to the cluster core. We also detect two new ultra-steep--spectrum ($\alpha \lesssim -1.7$) fossil plasma sources to the north and west of the cluster centre without associated compact radio emission. A previously detected radio galaxy in the cluster is also found to have a spectrally steepening tail with steep-spectrum components highlighted by the MWA. We do not detect a giant radio halo in the cluster. After simulating a range of radio haloes at 216~MHz we place upper limits on the radio luminosity \CORRS{corresponding to $\sim 7.2 \times 10^{24}$~W\,Hz$^{-1}$ at 150~MHz assuming the expected 500~kpc radius, up to a factor 5} lower than expected from scaling relations. \CORRS{Why a giant radio halo is undetected in Abell~3266 is unclear -- the timeline of the merger and overall mass of the system as determined by optical and X-ray studies suggest the foundation for a hosting system.}
%

\end{abstract}



\section{Introduction}

The observation of steep-spectrum ($\alpha \lesssim -1$ \footnote{We use the convention $S_\nu \propto \nu^\alpha$ for the radio spectral index, $\alpha$.}), diffuse synchtrotron radio emission in galaxy clusters provides a glimpse into the micro-Gauss--level magnetic fields \citep[e.g.][]{gf04} and particles of the intra-cluster medium (ICM). These radio sources are thought to relate to the physical processes that shape the largest scale structures of the Universe, generated through shocks and turbulence in the ICM \citep[for a review of diffuse radio sources in clusters, see][]{bj14,vda+19}.

The radio sources can be broken down into a handful of distinct, but observationally similar classes. These sources include megaparsec-scale, peripherally located, shock-driven radio relics (or `radio shocks', e.g.~\citealt{mj-h,Bagchi2006,vanWeeren2016,Duchesne2020b}), centrally located cluster merger and turbulence-driven giant, megaparsec-scale radio haloes \citep[e.g.][]{ffgg01,Wilber2020,vanWeeren2020,Duchesne2021a}, mini-haloes \citep[e.g.][]{Gitti2002,Giacintucci2014a,Duchesne2021b}, and a variety of other, generally smaller-scale re-accelerated/-energised radio plasmas \citep[e.g.][]{srm+01,vanWeeren2017,Mandal2020,Hodgson2021,Duchesne2021b}. Additionally, other aged -- `fossil' -- radio plasmas including remnant radio galaxies \citep[e.g.][]{mpm+11,Duchesne2020a,Quici2021} are observed in clusters. The link between these sources is not certain, and there is some suggestion of connection between the underlying particle populations and physical mechanisms as well as connections to active radio galaxies \citep[e.g.][]{Bonafede2014,vanWeeren2017,Jones2021,Duchesne2021b}. While diffuse cluster sources share similar observational properties, these sources differ fundamentally in the physical mechanisms that drive the acceleration of particles in the ICM magnetic fields even if their underlying particle populations originate from the same place. Complex merging cluster systems have been observed to host a combination of these diffuse radio sources \citep[e.g.][]{Pearce2017,Botteon2021,mgcls}, with a complex variety of merger dynamics and active radio galaxies within such systems \citep[e.g.][]{Breuer2020}. 

\subsection{Abell 3266}\label{sec:a3266}

Abell~3266 \citep{aco89} is a massive, merging galaxy cluster system located in the Horologium-Reticulum supercluster (HRS) \citep{fleenor05}, that contains several prominent radio-emitting objects. Over the past three decades Abell~3266 has been
the subject of numerous studies particularly in the optical and X-ray regimes \citep[e.g.][]{Robertson1990,Quintana1996,cypriano01,Finoguenov2006}. However, only minimal explorations have occurred at radio frequencies \citep[][and more recently by \citealt{Rudnick2021}]{murphy99, Miller2012, dehghan-phd, bvc+16, riseley-phd}, and these have been confined to frequencies above 800~MHz leaving the cluster largely unexplored at low radio frequencies ($\lesssim 400$~MHz).

{\citet{Finoguenov2006} noted that the cluster exhibits a low entropy region at its centre, but also that the entropy profile is significantly asymmetrical about the brightest cluster galaxy (BCG). While the low entropy region extends $\sim 8$~arcmin to the north-east, in the south-west the X-ray data show a sharp increase in entropy and temperature, bounded by a shock front (we refer the reader to fig.~1--3 in that paper).  It was concluded that this was the result of a sub-cluster group in-falling from the foreground, and in the plane of the sky. However, the X-ray images alone were insufficient to prove the existence of this sub-cluster companion.}

\citet{Dehghan2017} then explored the optical substructure of the system, finding five individual sub-clusters beyond the main cluster core at various stages of merging and dynamic activity. They report that the merging system is likely to have completed core passage of the main core components, and that a merger is well underway. Recently, \citet{Sanders2021} present \emph{eROSITA} \footnote{The \textit{extended ROentgen Survey with an Imaging Telescope Array} \citep{erosita2}.} X-ray observations of the system, further confirming the complex merger state and highlighting the numerous in-falling sub-clusters surrounding the central system.

{\citet{murphy99} reported re-processed observations of diffuse sources in the cluster at 843~MHz with the Molonglo Observatory Synthesis Telescope (MOST), taken as part of the Sydney University Molonglo Sky Survey \citep[SUMSS;][]{bls99}. These 843-MHz observations were also compared to higher frequency 1.4- and 2.4-GHz Australia Telescope Compact Array (ATCA) data. In particular, \citet{murphy99} noted the presence of a barely detectable ($2.5\sigma$) diffuse radio source to the south-east of the cluster in the 843-MHz data and not present in the higher frequency images. Given the low significance of this detection, the source was thought unlikely to be real. \citet{Miller2012} and \cite{dehghan-phd} considered 56 hours of available 1.4-GHz ATCA data of the cluster, but did not cover the region associated with the diffuse source reported by \cite{murphy99}, instead focusing on other compact sources in the region.}

\citet{bvc+16} and \citet{riseley-phd} observed the cluster with the KAT-7 \footnote{Karoo Array Telescope.} at 1.86~GHz and at 1.83 and 1.32~GHz, respectively. \citet{bvc+16} conclude the central emission in the cluster is composed of discrete radio sources and find no evidence of peripheral or other steep-spectrum emission that was not already revealed in the SUMSS image. \citet{riseley-phd}, however, notes there may be a hint of excess central diffuse emission after subtraction of contaminating discrete sources. The low resolution of the KAT-7 observations ($\gtrsim 3.5$~arcmin) limited the interpretation of this excess central emission. 

Abell~3266 was observed as part of the MeerKAT Galaxy Cluster Legacy Survey \citep[MGCLS;][]{mgcls}, where the candidate diffuse emission reported \citet{murphy99} was confirmed to be a peripheral relic. In addition to this,  \citet[][]{Rudnick2021} explored the smaller-scale structure of one of the cluster radio galaxies, highlighting its complex morphology with filamentary features and evidence of episodic activity.

In this work, we present new radio images of the Abell 3266 system from the Murchison Widefield Array \citep[MWA;][]{tgb+13} in its Phase II `extended' configuration \citep[][hereafter, MWA-2]{wtt+18}, providing the deepest, highest-resolution low-frequency view of this cluster to date. \citet{planck16} report a cluster mass of $6.64_{-0.12}^{+0.11} \times 10^{14}$~M$_\odot$~\footnote{Through the strength of the Sunyaev--Zeldovich (SZ) effect, though note a number of masses have been estimated for the cluster: $M_{500} = (4.11 \pm 0.96) \times 10^{14}$~M$_\odot$ \citep{spt2}, $M_{\text{X},500} \sim 4.6 \times 10^{14}$~M$_\odot$ \citep{pap+11}, $M_{500} = (8.80 \pm 0.57) \times 10^{14}$~M$_\odot$ \citep{Ettori2019}, through SZ, X-ray, and optical methods, respectively.} and \citet{Dehghan2017} report a mean cluster core redshift of $z = 0.0594 \pm 0.0005$. Throughout this paper, a standard $\Lambda$ Cold Dark Matter cosmology with $H_0 = 70$~km\,s$^{-1}$\,Mpc$^{-1}$, $\Omega_\text{M} = 0.3$, and $\Omega_\Lambda = 1-\Omega_\text{M}$ is assumed. At the redshift of Abell~3266 ($z=0.0594$) $1^{\prime}$ corresponds to 69~kpc.

\section{Data}

\begin{figure*}
\raggedright
\includegraphics[width=0.915\linewidth]{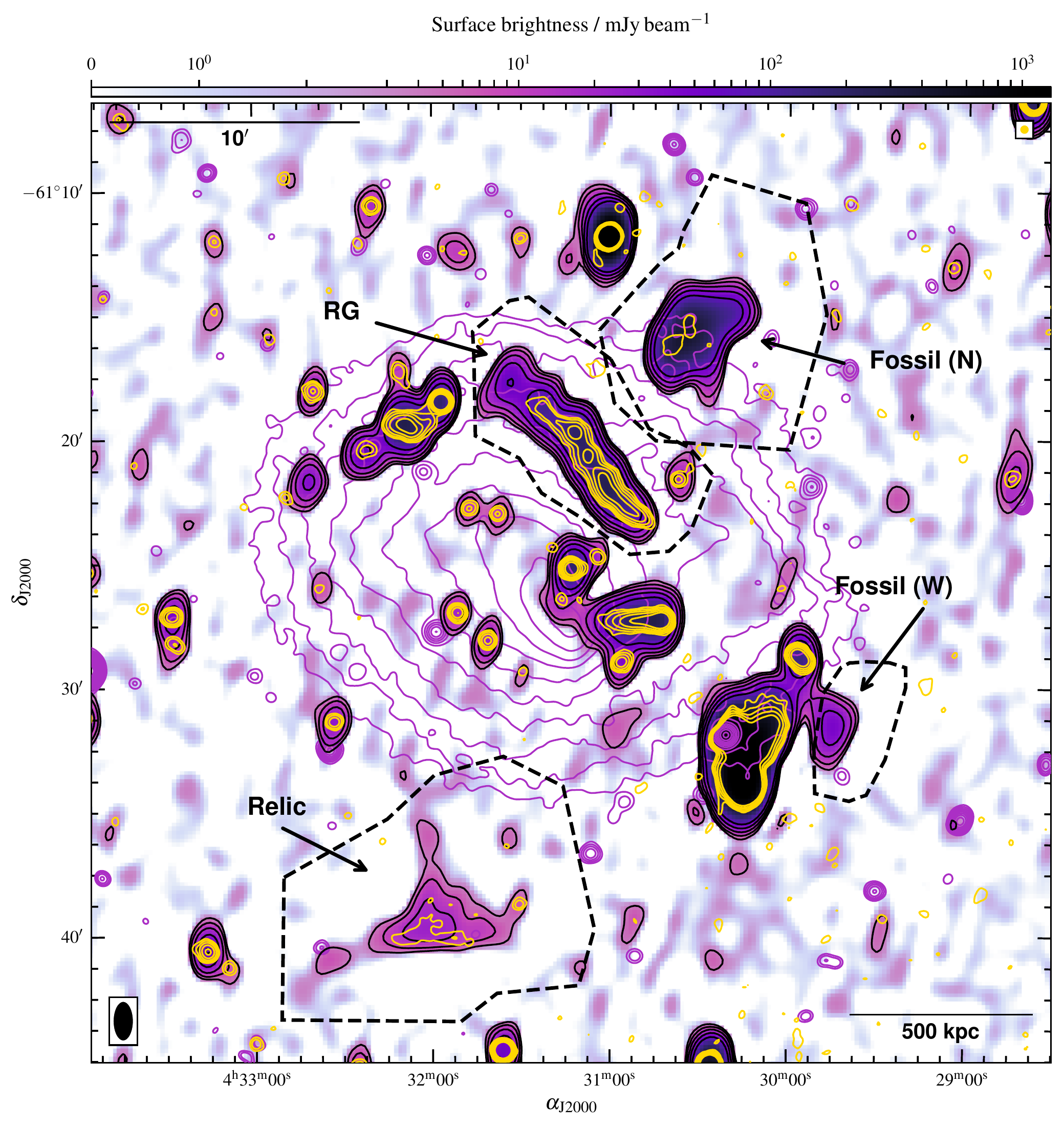}
\caption{\label{fig:a3266:mwa1} Abell~3266 at 216 MHz. The background image is the 216-MHz, robust $+0.5$ MWA-2 image and the black contours correspond to the background image with levels of $[3, 6, ..., 48]\times\sigma_\text{rms}$, where $\sigma_\text{rms} = 1.3$~mJy\,beam$^{-1}$. The overlaid contours in yellow are the re-imaged RACS robust $+0.25$ image at levels of $[3, 6, ..., 48]\times\sigma_\text{rms}$, where $\sigma_\text{rms} = 0.28$~mJy\,beam$^{-1}$. The purple contours are from the smoothed, exposure-corrected \xmm{} image. The sources of interest are labelled, and the black, dashed polygon regions are used for integrated flux density measurements. The framed ellipses in the lower left and top right are the shapes of restoring beams for the MWA-2 and RACS images, respectively. The linear scale is at the redshift of the cluster.}
\end{figure*}

\begin{table*}
    \centering
    \begin{threeparttable}
    \caption{\label{tab:mwaobs} Details of images used in this work.}
    \begin{tabular}{r c c c c c c c}
    \toprule
    $\nu$ \tnote{a} & Instrument &  $\tau_\text{effective}$ \tnote{b} & Image weighting & Restoring beam & min($\sigma_\text{rms}$) \tnote{c} & max($\theta$) \tnote{d} & $\xi_\text{scale}$ \tnote{e}  \\
        (MHz) & & (min) & & (arcsec $\times$ arcsec) & (mJy\,beam$^{-1}$)  & (arcmin) & (\%) \\\midrule
         88 & MWA-2 & 45 (48) & robust $+0.5$ ($+2.0$) & $197 \times 124$ ($260 \times 159$) & 8.3 (9.7 & 120 (60) & 10 (10)\\
         118 & MWA-2 & 57 (49) & robust $+0.5$ ($+2.0$) & $156 \times 92$ ($195 \times 117$) & 4.1 (6.0) & 120 (60) & 10 (9) \\
         154 & MWA-2 & 46 (56) & robust $+0.5$ ($+2.0$) &$116 \times 70$ ($149 \times 90$)  & 2.3 (4.0) & 120 (60) & 9 (9) \\
         185 & MWA-2 &30 (32) & robust $+0.5$ ($+2.0$) & $111 \times 59$ ($123 \times 76$)  & 2.8 (3.0) & 120 (60) & 9 (9) \\
         216 & MWA-2 & 81 (68) & robust $+0.5$ ($+2.0$) & $98 \times 51$ ($110 \times 72$) & 1.3 (1.5) & 120 (60) & 10 (10) \\\midrule
         200 \tnote{f} & MWA & $\sim 10$ & robust $-1.0$ & $149 \times 132$ & 22 &  $\sim 670$ & 8 \\
         887 & ASKAP & 15 & robust $+0.25$ & $25 \times 25$ \tnote{g} & 0.28 & 49 & 10 \\
         1372 \tnote{h} & KAT-7 & 270 & natural & $259 \times 218$ &  2.3 & 30.31 & 10 \\
         
    \bottomrule
    \end{tabular}
    \begin{tablenotes}[flushleft]
    {\footnotesize
     \item[] \textit{Notes.} For MWA-2 data, both robust $+0.5$ and (robust $+2.0$) images are used. \item[a] Central frequency. \item[b] Effective integration time. For the MWA-2 data this is the total integration time for all snapshots scaled by the stacked Stokes \textit{I} primary beam response at the position of Abell~3266 (i.e., less than the total integration time). \item[c] Minimum local rms noise near the cluster. \item[d] Maximum detectable angular scale. For MWA-2 robust $+2.0$ images this is lower as imaging with the $+2.0$ weighting becomes dominated by short baseline artefacts. This does not reduce sensitivity to scales of interest here by any signicant amount. \item[e] Uncertainty in the flux density scale. \item[f] See \citet{gleamegc} for imaging details. \item[g] Restored with a circular beam. \item[h] See \citet{riseley-phd} for imaging details.
    }
    \end{tablenotes}
    \end{threeparttable}
\end{table*}

This work is largely focused on low-frequency radio observations of Abell~3266, and makes use of data products from the MWA for characterisation of diffuse radio sources in the cluster. We supplement these data with the low-resolution 1372~MHz KAT-7 image published by \citet[][see their Chapter 3 for data reduction and imaging details]{riseley-phd}, and relevant observation and image details are shown in \autoref{tab:mwaobs} of this work.

\subsection{MWA}\label{sec:data:mwa}

The cluster has been observed by the MWA-2 in a 2-min snapshot mode, with observations centered on 88, 118, 154, 185, and 216~MHz with a 30-MHz bandwidth each. Due to the large field-of-view (FoV) of the MWA, snapshot observations within $\sim 5$ degrees are included for imaging in this work. At the lowest frequencies, more snapshots could be included due to the increase in FoV. However, at the lower frequencies we become more limited by confusion rather than sensitivity within and near the cluster. The MWA-2 fields `7' and `10' presented in \citet{Duchesne2021b} along with supplementary data from miscellaneous projects in the region are processed and co-added. Data processing follows the procedure outlined by \citet{Duchesne2020a,Duchesne2021a,Duchesne2021b} using the MWA calibration and imaging pipeline, \textsc{piip} \footnote{\url{https://gitlab.com/Sunmish/piip}}, and associated software. Namely, individual 2-min snapshots are calibrated and imaged independently prior to linear co-addition/mosaicking. Calibration takes place in-field with a local sky model generated from the GaLactic and Extragalactic All-sky MWA (GLEAM) survey \citep{wlb+15,gleamegc} and SUMSS \citep[see ][for details]{Duchesne2020a} and multi-scale, multi-frequency deconvolution is performed using \textsc{wsclean} \citep{wsclean1,wsclean2} with the \textsc{wgridder} algorithm \citep{wgridder1,wgridder2}. General astrometry and brightness scaling prior to co-addition is also as detailed in \citet{Duchesne2020a} and makes use of the aforementioned local sky model. This results in an overall flux scale uncertainty of $\sim 10$~per cent, predominantly inherited from the catalogues that comprise the sky model. 

At 216~MHz, with the smallest FoV the contribution from the more distant field `7' observations is minimal, however, the cluster has been observed more extensively at 216~MHz with a total integration time over all snapshots of 222~min. The low elevation of most of the individual snapshots, particularly at 216~MHz, results in lessened sensitivity across the main lobe of the primary beam. This limits the sensitivity in the final images, though we still obtain a root-mean-square (rms) noise in the stacked 216-MHz image of $\sim 1.3$~mJy\,beam$^{-1}$ near the cluster. All imaging details are shown in \autoref{tab:mwaobs}. 

We also make use of the 200-MHz wideband image from the GLEAM survey to complement the MWA-2 observations. Other GLEAM bands are generally less sensitive or are too confused for useful analysis in this case. 

\subsection{Other data products}

The Rapid ASKAP \footnote{Australian Square Kilometre Array Pathfinder \citep{Hotan2021}.} Continuum Survey (RACS; \citealt{racs1}, project AS110; \citealt{askap:racs}, SB8678) is also used in helping to identify contaminating discrete sources within the low-resolution MWA and KAT-7 data. Calibrated RACS visibilities from two adjacent observations covering the cluster are retrieved from the CSIRO \footnote{Commonwealth Scientific and Industrial Research Organisation.} ASKAP Science Data Archive \citep[CASDA;][]{casda,Huynh2020} and imaged with \textsc{wsclean}. We use a `Briggs' robust $+0.25$ image weighting following \citet{Duchesne2021b}, resulting in a $25^{\prime\prime} \times 25^{\prime\prime}$ resolution image with a local rms noise of $0.28$~mJy\,beam$^{-1}$ near the cluster. The brightness scale of this robust $+0.25$ image is consistent within a few per cent of the available RACS survey image for the region of interest. Further details of the RACS image are shown in \autoref{tab:mwaobs}.

For a qualitative overview of the cluster morphology, we use re-processed, archival \xmm{} data using the EPIC \footnote{European Photon Imaging Camera.} instrument \footnote{Obs.~IDs: 010526100, 0105262201, 0105260901, 0105260801, 0105262001, 0105262501, 0105260701, 0105261101, 0105262101.} (PI Aschenbach, originally presented by \citealt{Sauvageot2005,Finoguenov2006}). For detail about the \xmm{} data reduction, we refer to appendix~A of \citet[][]{bartalucci2017}. 


\section{Results and discussion}

\begin{figure*}
\includegraphics[width=1\linewidth]{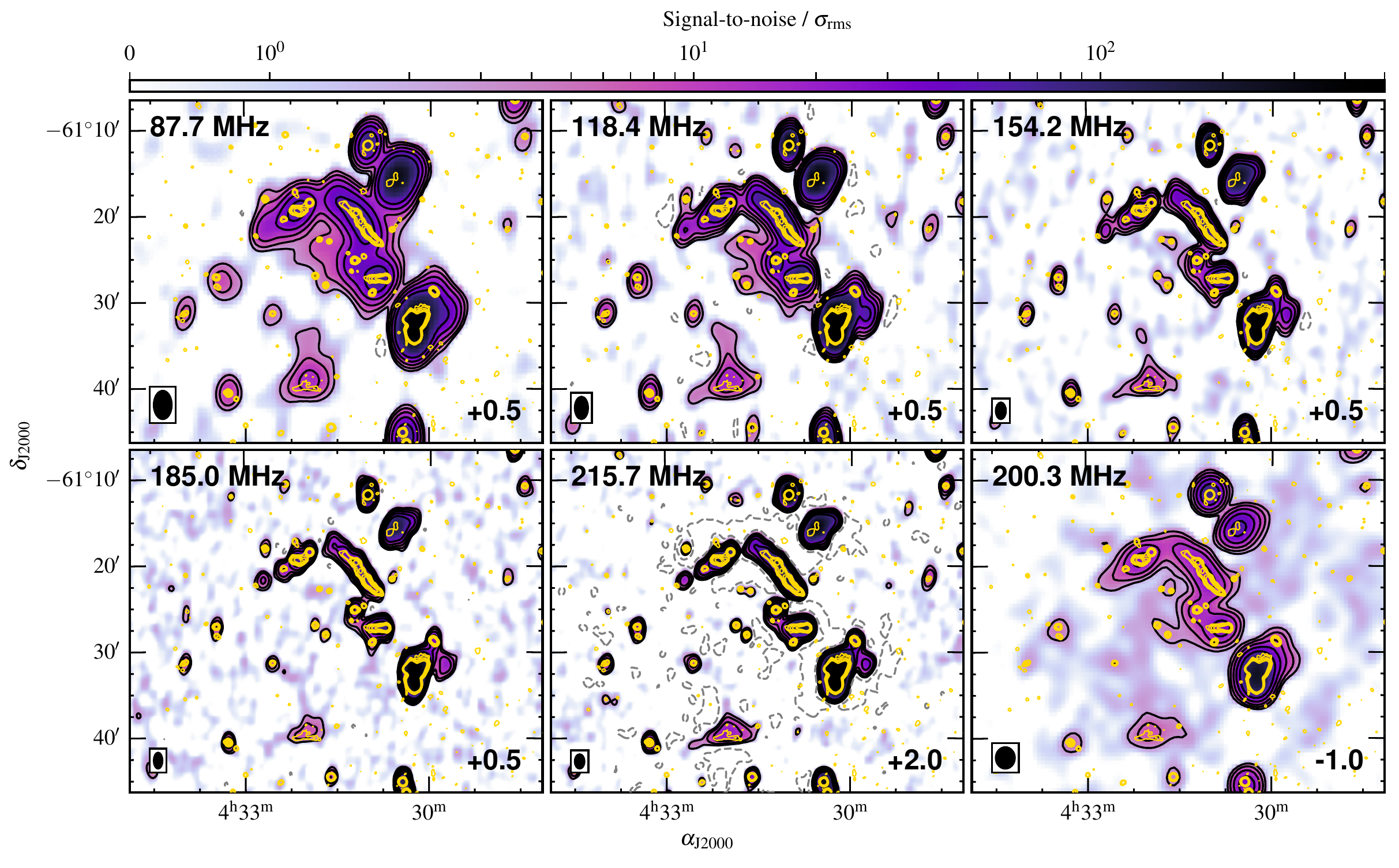}
\caption{\label{fig:a3266:all} MWA images of Abell~3266 from 88--216~MHz. The `Briggs' robust parameter for each image is labelled. The MWA-2, 216-MHz robust $+0.5$ image is shown in \autoref{fig:a3266:mwa1}. Black contours are drawn at $[3, 6, 12, 24, 48]\times\sigma_\text{rms}$, with $\sigma_\text{rms}$ for each image reported in \autoref{tab:mwaobs}\CORRS{, and the grey, dashed contour is set at $-3\sigma_\text{rms}$}. Yellow contours are from the RACS image as in \autoref{fig:a3266:mwa1}. Note the colourscale for all images is mapped to 0--500$\sigma_\text{rms}$.}
\end{figure*}

\begin{figure}
    \centering
    \includegraphics[width=1\linewidth]{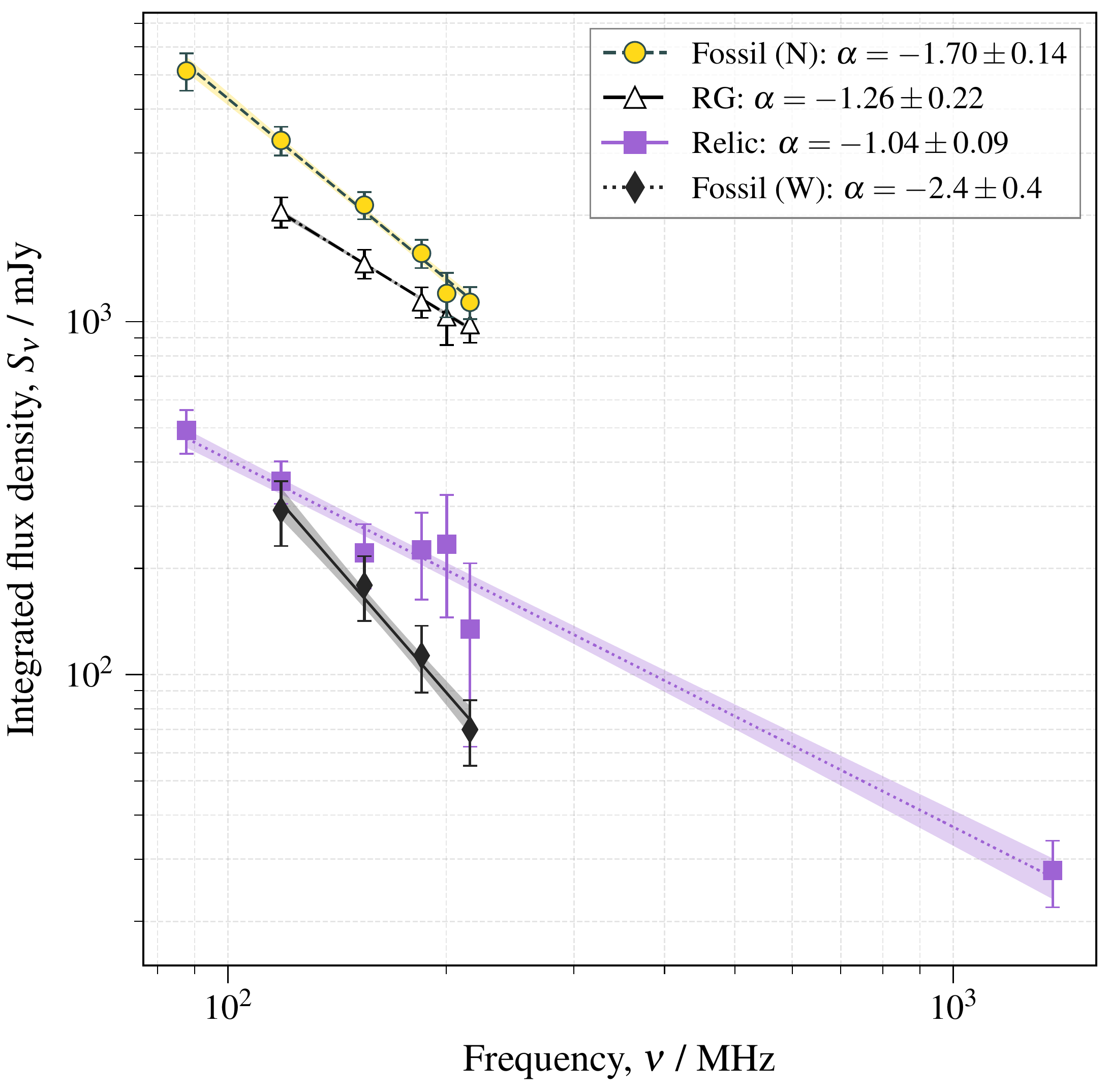}
    \caption{\label{fig:seds} Integrated spectra of the four radio sources of interest. Measurements are also provided in \autoref{tab:flux}. The sources are fit with normal power law models, and a 68\% confidence interval for each fit is shown. }
\end{figure}

\begin{table}
    \centering
    \begin{threeparttable}
    \caption{\label{tab:flux} MWA-2, GLEAM, and KAT-7 flux density ($S_\nu$) measurements of the four sources labelled on \autoref{fig:a3266:mwa1}, integrated within the regions shown on \autoref{fig:a3266:mwa1}.}
    \begin{tabular}{r c c c}\toprule
         $\nu$ & $S_\nu$  & $S_{\nu,\text{discrete}}$ \tnote{a} & $\alpha$ \\
          (MHz) & (mJy) & (mJy) & \\\midrule
          \multicolumn{4}{c}{Fossil (N)} \\\midrule
87.7 & $5130 \pm 620$ & - & \multirow{6}{*}{$-1.70 \pm 0.14$} \\
118.4 & $3260 \pm 310$ & -&\\
154.2 & $2140 \pm 190$ & -&\\
185.0 & $1560 \pm 140$ & -&\\
200.3 & $1110 \pm 190$ & -& \\
215.7 & $1130 \pm 120$ & -&\\
\midrule
          \multicolumn{4}{c}{RG} \\\midrule
118.4 & $2050 \pm 200$ & $30 \pm 3$ & \multirow{5}{*}{$-1.26 \pm 0.22$} \\
154.2 & $1460 \pm 140$ & $ 22 \pm 2$ & \\
185.0 & $1140 \pm 110$ & $19 \pm 2 $ &\\
200.3 & $1030 \pm 170$ & $18 \pm 1 $ & \\
215.7 & $980 \pm 110$ & $16 \pm 1 $ &\\\midrule
          \multicolumn{4}{c}{Relic} \\\midrule
87.7 & $492 \pm 69$ & $21 \pm 15$ & \multirow{6}{*}{$-1.04 \pm 0.09$}\\
118.4 & $353 \pm 48$ & $17 \pm 10$ &\\
154.2 & $221 \pm 46$ & $14 \pm 7$&\\
185.0 & $225 \pm 62$ & $13 \pm 5$&\\
200.3 & $211 \pm 92$ & $12 \pm 5$&\\
215.7 & $135 \pm 72$ & $11 \pm 4$&\\
1372.0 & $28.9 \pm 6.0$ & $3.1 \pm 0.3$ & \\\midrule
          \multicolumn{4}{c}{Fossil (W)} \\\midrule
118.4 & $292 \pm 61$ & - &\multirow{4}{*}{$-2.4 \pm 0.4$}\\
154.2 & $179 \pm 37$ &-&\\
185.0 & $113 \pm 24$ & -&\\
215.7 & $70 \pm 15$ &-&\\
\bottomrule
    \end{tabular}
    \begin{tablenotes}[flushleft]
    {\footnotesize \item[a] Subtracted contribution from discrete sources detected in the RACS image, assuming $\alpha = -0.7$.}
    \end{tablenotes}
    \end{threeparttable}    
\end{table}

\subsection{The diffuse radio sources}

\autoref{fig:a3266:mwa1} shows the cluster at 216~MHz with the robust $+0.5$ MWA-2 data along with the re-imaged RACS image overlaid as contours. The \xmm{} image is shown as purple contours highlighting the complex morphology of the ICM of the system. These data reveal numerous radio sources, and four sources are highlighted and labelled on \autoref{fig:a3266:mwa1}, including the bright, elongated radio galaxy north of the cluster centre (`RG'), what we consider a relic towards the south (`Relic'), a fossil plasma source towards the north (`Fossil (N)'), and a smaller fossil plasma source nearby the bright, complex head-tail galaxy towards the west (`Fossil (W)'). 

\autoref{fig:a3266:all} shows additional 88--216~MHz MWA images of the cluster, with most features present in all images. Flux densities, $S_\nu$, for all sources are integrated within the polygon regions shown on \autoref{fig:a3266:mwa1} following the process described in Section 2.4 of \citet{Duchesne2021b}. Uncertainties for flux density measurements, $\sigma_{S_\nu}$, are \begin{equation}\label{eq:uncertainty}
    {\sigma_{S_\nu}}^2 = N_\text{beam}{\sigma_\text{rms}}^2 + \left(\xi_\text{scale} S_\nu\right)^2 + {\sigma_\text{discrete}}^2 \, ,
\end{equation}
where $N_\text{beam}$ is the number of independent restoring beams covering the integration region, $\xi_\text{scale}$ is the flux scale uncertainty ($\sim 10$~per cent for most images), and $\sigma_\text{discrete}$ is the uncertainty on the flux density for subtracted discrete sources. Relevant discrete sources detected in the RACS image are subtracted by extrapolating their 887-MHz flux density to other frequencies assuming a power law model with spectral index $\alpha = -0.7$. \CORRS{Additionally, $\sigma_\text{discrete}$ in \autoref{eq:uncertainty} incorporates both the measurement uncertainty of the discrete sources and the uncertainty in this assumed $\alpha$ by adding in quadrature the difference between the extrapolated flux densities when using a range of $\alpha$ \citep[between $-0.5$ and $-1.0$; for further detail refer to Section 2.4 of][]{Duchesne2021b}.} We use the robust $+2.0$ MWA-2 images for measuring the flux densities of all sources except the western fossil, which becomes too blended with the nearby wide-angle tail (WAT) source, PKS~0429$-$61. For the western fossil, we instead use the robust $+0.5$ MWA-2 images, convolving all maps to a common resolution (that of the 118-MHz robust $+0.5$ map) to ensure any additional blending with the WAT is consistent across the band.

We note that the higher background signal in the GLEAM 200-MHz image results in a large measurement uncertainty. Individual flux density measurements, along with subtracted discrete source contributions, are reported in \autoref{tab:flux}. All four sources are fit with normal power law models, and the measured spectra and best-fitting models are shown on \autoref{fig:seds}.

\subsubsection{Radio galaxy (RG)}

To the north of the cluster centre, an elongated, 630~kpc radio galaxy is observed, detected completely across the MWA data as well as in SUMSS and the RACS image. This source is also detected in the ATCA data presented by \citet{murphy99}, who note a possible optical host at the south-west `head' and a spectral steepening along the `tail'. As seen in \autoref{fig:a3266:mwa1}, the RG is almost completely detected in the RACS image, except for the steepest-spectrum component at the `tail' end to the NE. Observationally, the source is similar to the large head-tail radio galaxy in Abell~1132 \citep[][]{Wilber2018}. The integrated spectral index of $\alpha = -1.26 \pm 0.22$ across the MWA band highlights the contribution from the steep-spectrum components along the `tail' of the source.

\citet{Rudnick2021} present higher-resolution MeerKAT data at 1.285~GHz for this source, showing a complex filamentary morphology with `ribs' and `tethers', not discernible in the lower-resolution data presented here except as some clumping along the emission. The extent of the reported emission is similar to that seen in the RACS data. They report a faint core towards the south-west (i.e. the `head' of the emission), associated with a spectroscopically-confirmed cluster member. \citet{Rudnick2021} show a resolved spectral index map for the source across the 900--1670~MHz MeerKAT data, highlighting this steepening from $\alpha \approx -0.6$ at the south-west `head' of the radio emission to $\alpha \lesssim -3$ approaching the base of the north-east `tail'. In the 216-MHz MWA-2 data shown in \autoref{fig:a3266:mwa1}, the additional component at the end of the north-east `tail' is not detected in the high-resolution MeerKAT image. 

This additional component of the tail is not measured separately in this work due to significant confusion, and a resolved spectral index map is not possible with the present MWA data. Based on the integrated spectral index we report here ($\alpha = -1.26 \pm 0.22$) across the entire structure, observation of such steep-spectrum components undetected in the MeerKAT image implies some curvature beyond $\sim 216$~MHz.

\subsubsection{Relic}

\citet{murphy99} reported a 2.5$\sigma_\text{rms}$ detection of a candidate relic to the south-west of the cluster, which is confirmed by the 1.285-GHz MeerKAT data published by \citet{mgcls}. The MWA and KAT-7 data additionally detect this relic source. The (deconvolved) largest angular size is $\sim 8.3$ arcmin in the 216-MHz image, corresponding to $\sim 570$~kpc, smaller than the size reported by \citet{mgcls} \footnote{They report 9.7~arcmin, though after inspection of the publicly available MGCLS images we measure a largest angular extent of $\sim 8.6$~arcmin, closer to our MWA measurement.}. The integrated spectrum between 88--1327~MHz has a spectral index of $\alpha = -1.04 \pm0.09$, consistent with the general relic population and as expected from diffusive-shock acceleration processes. The relic is also partially detected in the RACS image, though the emission is patchy and we note that the RACS observations have low sensitivity to angular scale approaching 10~arcmin \citep[][]{racs1,Duchesne2021b}. 

As we decrease in frequency through the MWA-2 data, the relic is shown to extend further core-ward with a `bridge' of emission extending for $\sim 4.5$~arcmin ($\sim 310$~kpc) from the relic to the cluster core. No discrete sources are detected in the RACS image at this location, and the lack of detection in the 216-MHz MWA-2 image suggests spectral steepening in this direction. Such extensions connected to radio relics have sometimes been observed to connect to giant radio haloes, and typically show a spectral gradient steepening in the core-ward direction indicative of an ageing radio plasma \citep[e.g.][]{vanWeeren2016}.

\subsubsection{Fossil (N)}

To the north of the cluster we detect a previously unreported steep-spectrum source which we consider fossil radio plasma. The source has a slightly elongated morphology with an angular size of $\sim 5.4$~arcmin ($\sim 370$~kpc) and is located in the space between the spectroscopic groups `3', `4', and `5'  from \citet[][see also fig.~6.6 from \citealt{dehghan-phd}]{Dehghan2017}. No compact emission is seen in the RACS data within the emission region at 216~MHz. As with the Southern relic, the RACS image shows a patchy detection. We find $\alpha = -1.70 \pm 0.14$ across the available MWA data, and consider this another example of cluster-based fossil plasma from a long-dead active galactic nucleus (AGN). 

Alternatively, the source may represent a radio phoenix or other re-accelerated fossil plasma source. The cluster is sufficiently dynamic that weak shocks will be present throughout the volume and indeed \citet{Sanders2021} report two candidate shocks, though not at this location. While not reported by \citet{mgcls}, the source is also detected in the publicly available images from the MGCLS. The MGCLS data as well as deeper ASKAP observations will provide high frequency data points to constrain the spectrum and investigate relevant synchrotron ageing models to determine the age of the source (e.g., Riseley et al., in prep.).

\subsubsection{Fossil (W)}

The fossil source to the west of the complex WAT has an ultra-steep spectrum with $\alpha = -2.4\pm0.4$ between 118--216~MHz. This source is smaller than other diffuse sources in the cluster at $\sim 2.4$~arcmin ($\sim 170$~kpc). It may represent a past episode from the WAT, however, it appears to be a distinct component, where the blending between the WAT and fossil is a function of image resolution rather than physical proximity. As with the northern fossil source this may also instead represent a re-accelerated plasma, though with the present data it is not possible to distinguish between a remnant or re-accelerated radio plasma.  

 Such ultra-steep--spectrum emission has been steadily uncovered in MWA data \citep[][]{Hodgson2021,Duchesne2021b,Duchesne2017} and highlights a niche discovery space for the MWA and other low-frequency inteferometers. However, with the low resolution and low signal-to-noise ratio (SNR) in these detections, uncertainties remain in the integrated spectra for these faint ultra-steep--spectrum radio sources detected by the MWA, and indeed it will not be until SKA-Low begins operations that we will be able to explore them in much more detail at these low frequencies.

\subsection{The absence of a giant radio halo in Abell~3266}\label{sec:halo}

\begin{figure*}
\includegraphics[width=1\linewidth]{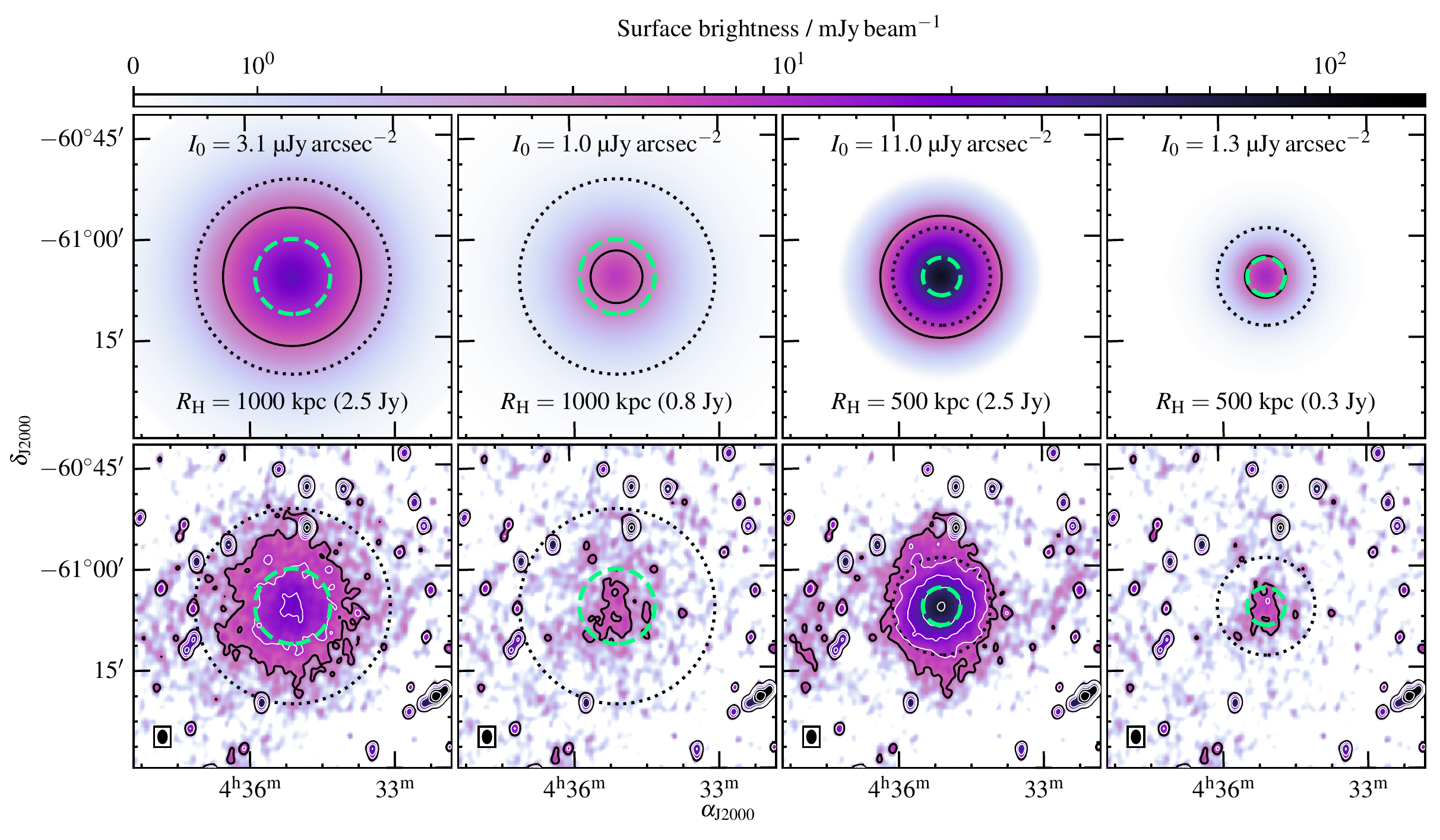}
\includegraphics[width=1\linewidth]{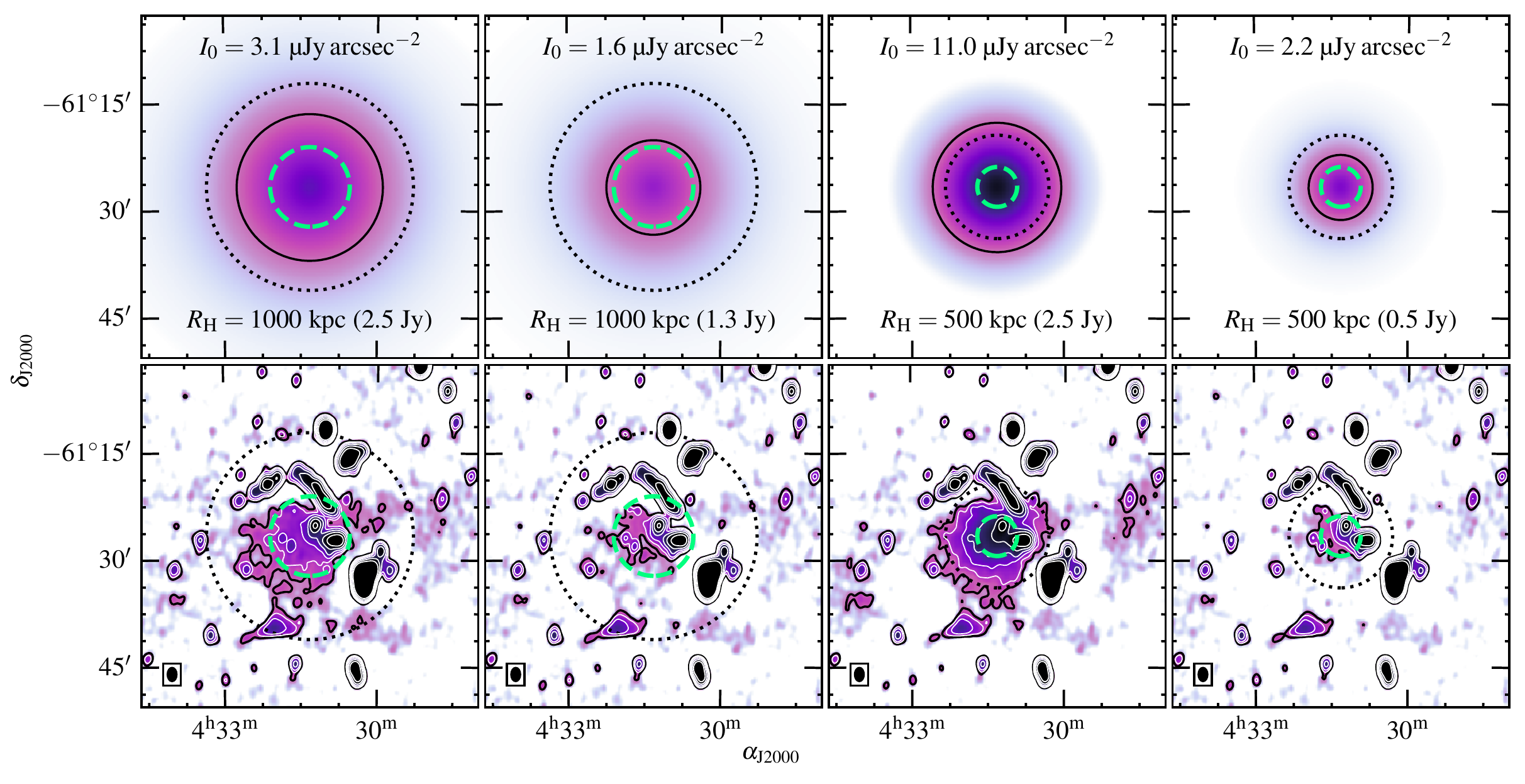}
\caption{\CORRS{\label{fig:a3266:mock} Mock radio haloes injected into a quiet patch of sky near Abell~3266 (\textit{top two rows}) and at the cluster centre(\textit{bottom two rows}). The model radio haloes are convolved with a  $100^{\prime\prime} \times 70^{\prime\prime}$ beam for comparison and are shown prior to imaging and injection into the datasets. The quoted surface brightness in each panel is the peak value at 216~MHz after injecting into the datasets. The solid contour is set at $3\sigma_\text{rms}$ ($\sigma_\text{rms} = 1.5$~mJy\,beam$^{-1}$, corresponding to $0.19$~\textmu Jy\,arcsec$^{-2}$), and the solid white contours correspond to $[6, 12, 24, 48, 96]\times\sigma_\text{rms}$. The dotted circles indicate the $R_\text{H}$ for each model, and the smaller dashed, green circles are $r_e = R_\text{H}/2.6$. The haloes predicted from the $P_{150~\text{MHz}}$--$M$ scaling relation (with $S_{216} = 2.5$~Jy, assuming $\alpha = -1.4$) and the upper limits are shown for $R_\text{H}=1000$~kpc (14.4~arcmin) and $R_\text{H}=500$~kpc (7.2~arcmin). For the top panel, Abell~3266 is featured in the lower right corner. The colour scale is the same in all panels and all panels have the same angular scale. Note the 216-MHz robust $+2.0$ image with no simulated halo is shown in \autoref{fig:a3266:all}.}}

\end{figure*}

Many massive ($\gtrsim 5 \times 10^{14}$~M$_\odot$), merging clusters have been observed to host giant ($\gtrsim 1$~Mpc) radio haloes \citep[e.g.][and references therein]{vda+19}, with luminosities that have been found to scale with the cluster mass and other physical properties \citep[e.g.][]{lhba00,Cassano2010,mfj+16} at both 1.4~GHz \citep[e.g.][]{ceb+13,Duchesne2017,Cuciti2021b} and 150~MHz \citep[][]{vanWeeren2020,Duchesne2021a}.

While Abell~3266 hosts a mix of extended, steep-spectrum radio sources, MWA-2 data supports the lack of a central giant radio halo as noted by \citet{bvc+16}. The MGCLS image reveal some diffuse emission connected to a central tailed galaxy \citep[considered a `confused tail' by][]{mgcls}, though the emission is largely enclosed within the $3\sigma_\text{rms}$ contour of the MWA-2 216-MHz image (e.g. \autoref{fig:a3266:mwa1}) around the radio sources near the cluster centre. This emission will be explored further by Riseley et al. (in prep.) but we do not consider this to be a giant radio halo.

We note the GLEAM 200-MHz image shows a high background signal around the cluster (see the bottom right panel of \autoref{fig:a3266:all}), however, this is largely from \CORRS{residual} sidelobes from the snapshot imaging process and is a feature of bright, extended sources throughout the GLEAM survey. This is particularly noticeable in the 200-MHz wideband images. 
Significant source confusion occurs in the lower two MWA-2 bands near the cluster centre. It is not clear from the images at 88- and 118-MHz that there is any excess diffuse emission beyond contributions from (1) discrete sources, (2) the previously noted diffuse emission, and (3) the additional ionospheric blurring that more heavily affects the 88- and 118-MHz data. The less confused, higher-sensitivity 216-MHz images do not reveal excess emission throughout the cluster that would hint towards a giant radio halo.

Based on the non-detection in the deep 216-MHz data, we investigate two possibilities: (1) a halo exists at the expected brightness based on established power--mass scaling relations \citep[e.g.][]{basu12,ceb+13} and is undetected in the 216-MHz \footnote{We choose the 216-MHz data due to the much longer integration time compared to the other bands combined with the highest nominal resolution resulting in less confusion.} MWA-2 data due to limited sensitivity to large angular scales, or (2) a halo, if present, is under-luminous with respect to the scaling relation. Such under-luminous haloes are being uncovered in high-sensitivity observations both in high- and low-resolution imaging. An example is the radio halo in Abell~3667: first detected in single-dish imaging from by the Parkes radio telescope \citep{Carretti2013}, followed up in low-resolution MWA-1 data \citep{Hindson2014} and KAT-7 imaging \citep{Riseley2015}, and finally confirmed by recent MGCLS observations \citep{deGasperin2021b} to be a giant, under-luminous radio halo \footnote{Note a hint of emission had been reported by \citet{mj-h}, though it was unclear at the time if the emission was made up of artefacts from the nearby bright source.}.

\subsubsection{Simulating a giant radio halo}

To simulate the giant radio halo, we estimate the expected halo brightness based on the scaling relation derived by \citet[][]{Duchesne2021a}, between cluster mass within $r_{500}$ \footnote{The radius within which the mean density of the cluster is 500 times the critical density of the Universe.} ($M_{500}$) and radio luminosity at 150~MHz ($P_{\text{150~MHz}}$) \CORRS{including the full literature sample of radio haloes.} \CORRS{We note \citet{vanWeeren2020} originally derived the 150-MHz scaling relation from a sample of radio haloes detected at frequencies close to 150~MHz. This may bias the sample towards a larger fraction of ultra-steep--spectrum radio haloes \citep[with $\alpha \lesssim -1.5$, hereinafter USSRH;][]{bgc+08} which are found to be under-luminous with respect to scaling relations derived at 1.4~GHz \citep[][]{ceb+13}.}

For consistency with \CORRS{scaling relations derived by \citet{Duchesne2021a}}, we use the PSZ2 mass of $\sim 6.64 \times 10^{14}$~M$_\odot$. The expected power at 150~MHz is $\sim 3.5 \times 10^{25}$~W\,Hz$^{-1}$, corresponding to a flux density of $\sim 4.2$~Jy, or $\sim 2.5$~Jy at 216~MHz, assuming a spectral index of $-1.4$ \footnote{For $\alpha > -1.4$, the brightness increases at 216~MHz and the halo will become more easily detectable.} \citep[as in][]{Duchesne2021a}. The radial size of radio haloes has been found to somewhat correlate to their luminosity and host cluster properties \citep[][]{Feretti2000,Cassano2007} and following the relation obtained by \citet{Cassano2007} we would expect an average radio halo radius of $\sim 450$~kpc. The correlation has recently been updated by \citet{Hodgson2021b} with radio halo sizes compiled by \citet[][and references therein]{sjp16}, though they note the correlation is poor we obtain a radius of $\sim 500$~kpc. As a point of comparison, we also consider a larger halo at $R_\text{H}=1000$~kpc which serves to both ensure a suitable limit is estimated if the radius is large but also accounts for a generally smoother brightness profile. We consider a limiting case where patchy, extended emission is seen in the final image which would indicate a detection, \CORRS{noting that $\sigma_\text{rms} \simeq 1.5~\text{mJy}\,\text{beam}^{-1} \simeq 0.19~\text{\textmu Jy}\,\text{arcsec}^{-2}$ for the 216-MHz robust~$+2.0$ imaging.}

\CORRS{Assuming $\alpha = -1.4$, frequency-dependent model} radio haloes are simulated in each 216-MHz snapshot independently following the standard circular exponential profile \citep[e.g.][]{Orru2007,Murgia2009,Bonafede2009,Boxelaar2021}, \begin{equation}\label{eq:profile}
I(r) = I_0 \exp\left( -r/r_e\right) \quad [\text{Jy\,arcsec}^{-2}] \, , 
\end{equation} with the peak surface brightness $I_0$, and the $e$-folding radius $r_e$ which is typically found to be $\sim R_\text{H} / 2.6$ \citep{Bonafede2017} for a radio halo of radius $R_\text{H}$. \CORRS{We simulated haloes in two separate locations: at the cluster centre as defined by the coordinates from the Meta-Catalogue of X-ray detected Clusters of galaxies \citep[MCXC;][]{pap+11}, and in a relatively empty patch of sky near the cluster. Each location is tested independently. } 

\CORRS{To avoid the somewhat uncertain primary beam attenuation \footnote{\CORRS{Particularly the difference in linear instrumental polarizations, XX and YY, which can vary significantly for off-zenith observations and differ from the tile-average model beam \citep{Chokshi2021}.}} the model haloes are Fourier transformed into the datasets and imaged separately down to the same CLEAN threshold and with identical $u,v$ sampling and image and visibility weighting as the normal data. These snapshot-dependent model-only images are then linearly added to the normal snapshot images, allowing us to avoid the complex primary beam attenuation.}

The \CORRS{combined model and data} snapshots are then mosaicked in the usual way, though we forgo \CORRS{further} brightness scaling and astrometry corrections. The ionospheric blurring is minimal at 216~MHz and the flux scale deviation between snapshots is small \footnote{The snapshots are on average $\sim 15$~per cent brighter than expected, which is corrected during normal imaging as discussed in Section~\ref{sec:data:mwa}.} and less important here as only the SNR matters for a detection. The total flux density is obtained from the input model rather than measured in the image. \CORRS{The} $R_\text{H} = 1000~\text{kpc} = 14.4~\text{arcmin}$ case illustrates the limitation of the MWA-2 at 216~MHz where it is clear the recovery of the simulated halo in the stacked image is incomplete by up to a factor of $\sim 5$ due to the low SNR at this large angular scale.


\begin{table}
    \centering
    \begin{threeparttable}
    \caption{\label{tab:limits} Derived upper limits for a giant radio halo in Abell~3266 represented by \autoref{eq:profile}.}
    \begin{tabular}{c c c c c }\toprule
         $\nu$ & $S_\nu^{500~\text{kpc}}$ & $S_\nu^{1~\text{Mpc}}$ & $P_{\nu}^{500~\text{kpc}}$ & $P_{\nu}^{1~\text{Mpc}}$ \\
         (MHz) & \multicolumn{2}{c}{(mJy)}  & \multicolumn{2}{c}{($\times
         10^{24}$~W\,Hz$^{-1}$)} \\\midrule
         \multicolumn{5}{c}{Cluster centre}\\\midrule
         150 \tnote{a} & $<800$  & $<2200$ & $<7.2$ & $<18$ \\
         216 & $<500$ & $<1300$ & $<4.3$ & $<11$  \\
         1400 \tnote{a} & $<40$ & $<90$ & $<0.3$ & $<0.8$ \\\midrule
         \multicolumn{5}{c}{Quiet sky}\\\midrule
         150 \tnote{a} & $<500$  & $<1300$ & $<4.3$ & $<12$ \\
         216 & $<300$ & $<800$ & $<2.6$ & $<6.9$  \\
         1400 \tnote{a} & $<20$ & $<60$ & $<0.2$ & $<0.5$ \\\bottomrule
    \end{tabular}
    \begin{tablenotes}[flushleft]
    {\footnotesize \item[a] Scaled from limits derived at 216~MHz assuming $\alpha = -1.4$.}
    \end{tablenotes}
    \end{threeparttable} 
\end{table}

The resultant models and their detection in the stacked 216-MHz images are shown in \autoref{fig:a3266:mock}, with the expected flux density modelled (2.5~Jy) along with what we find to be upper limits \CORRS{for the quiet sky case (0.8~Jy and 0.3~Jy at 216~MHz for radii of $R_\text{H}=1000$~kpc and $R_\text{H}=500$~kpc, respectively) and the cluster centrecase (1.3~Jy and 0.5~Jy)}. \CORRS{A slightly higher limit is obtained for the cluster centredue to the residual negative sidelobes around some of the central sources which is a result of the limited deconvolution in individual snapshots.}

\CORRS{Corresponding luminosities at 150~MHz, 216~MHz, and 1.4~GHz along with derived flux densities are reported in \autoref{tab:limits}.}

\subsubsection{\CORRS{Why do we not detect a giant radio halo?}}

\CORRS{The derived limits place the expected $R_\text{H} = 500$~kpc radio halo to be $\sim 5$ times below the 150-MHz scaling relations reported by \citet{Duchesne2021a}, though only $\sim 2.5$ times below the scaling relations reported by \citet{vanWeeren2020}. If the halo were to be offset slightly from the cluster X-ray center, the limit decreases to approach the quiet sky limit, up to $\sim 8$ times below the scaling relations. Conversely, increasing the halo size to $R_\text{H}=1000$~kpc provides a limit approximately in line with the \citet{vanWeeren2020} results and up $\sim 2$ times below the \citet{Duchesne2021a} relations. Historically, such large haloes are rarely observed \citep[see e.g. halo radii compiled by][]{Cassano2007,sjp16,Hodgson2021b} so this scenario is less likely.}

Mergers are expected to trigger radio haloes in massive clusters \citep[e.g][]{ceb+13}, however, a number of merging cluster systems are absent of radio haloes and upper limits below the established scaling relations have been placed \citep[e.g.][]{Cassano2016,Bonafede2017}. Additionally, some giant radio haloes have be observed to be under-luminous with respect to the 1.4~GHz scaling relations \citep[e.g. by a factor of $\sim 5$;][]{Cuciti2018}. Such discrepancies are usually \CORRS{USSRH \citep[e.g.][]{bgc+08,ceb+13,vanWeeren2020}} or arise due to systematic measurement-related problems. Incomplete $u,v$ sampling results in a loss of flux on large angular scales \citep[e.g.][]{Cuciti2018}. Measurements with a low SNR can also under-estimate integrated flux densities which can be mitigated somewhat by fitting a model to the surface brightness profile \citep[e.g.][]{Bonafede2017,Boxelaar2021}. In this work the flux density limits are obtained from the input model rather than measured in the resultant image \CORRS{avoiding this issue}, and limits are derived from the 150-MHz scaling relations where USSRH are not under-luminous \citep[e.g.][]{Duchesne2021a}. We have assumed $\alpha = -1.4$ for our simulated radio haloes, however, if we consider $\alpha=-1$ the expected halo brightness at 216~MHz increases. Conversely, $\alpha < -4$ is required to push a $S_\text{150~MHz} = 4.2$~Jy radio halo beyond detection in the 216-MHz image, hence, we do not consider that an ultra-steep--spectrum is responsible for the non-detection. 

We instead consider the physical picture of the Abell~3266 system that may limit generation and/or observation of a giant radio halo. The observable lifetimes of haloes may be shorter than the merger processes \CORRS{\citep[e.g.][]{ddbc13,Cassano2016}} and cluster systems in a pre-merging state or at a late stage in their merger may not be observed with radio haloes as there may not be sufficient turbulence within the cluster volume to generate the radio emission \citep{Bonafede2017}. These pre-merging or late-merger systems may also be expected to host ultra-steep spectrum radio haloes \citep[$\alpha \lesssim -1.5$;][]{bgc+08}, though no reasonable spectral index would shift the expected brightness at 216~MHz beyond detectability in this particular case. Some pre-merging systems have been found to host radio haloes \citep[e.g.][]{Ogrean2015,Duchesne2021a}, and a halo is sometimes found in individual sub-clusters \citep{Murgia2010,Botteon2018} \footnote{In these particular systems radio bridges have also been found between sub-clusters.}.

\textit{`Pre-merging', a late merger, or a weak merger?} \citet{dehghan-phd} investigated the evolution of the merger for the cluster using a Newtonian $N$-body simulation between the main two optical substructure components of the core region. The result suggested the distribution of galaxies observed represents the system $\sim 0.2$--$0.3$~Gyr after core-passage\CORRS{, predominantly involving two sub-clusters with a mass ratio of 1:2 \citep[furthermore see][]{Dehghan2017}.} \citet{dehghan-phd} notes this is consistent with both a low-entropy tail reported by \citet{Finoguenov2006} as well as the complex radio structure in the bright WAT. Such a scenario suggests the system is not in a pre-merging state. Furthermore, the \textit{eROSITA} entropy map presented by \citet{Sanders2021} support the passage of the north-east sub-cluster through the core region which has also been suggested by \citet{Sauvageot2005}, after core passage $\sim 0.2$~Gyr ago.The system is therefore unlikely to be in a pre-merging state. Weak mergers may produce fainter, perhaps undetectable haloes in part due to steeper spectra \citep[e.g.][]{bgc+08,Cassano2010}. A less likely scenario proposed by \citet{Sauvageot2005} suggests the the merger state is $\sim 0.8$~Gyr after core passage, however, a major merger timescale is expected to be $\sim 1$--3~Gyr from simulations \citep[e.g.][]{Cassano2016}. The optical substructure analysis presented by \citet{Dehghan2017} highlights the highly-energetic, complex, multi-component merger depositing large amounts of energy into the ICM \CORRS{with a mass ratio of 1:2 for the dominant central components. Such major mergers are likely a requirement for the formation of giant radio haloes \citep[e.g.][]{Cassano2016}.} Candidate shocks reported by \citet{Sanders2021} and the detection of a radio relic \CORRS{may further support} this \CORRS{major merger scenario, though \citet{Sanders2021} note it is not yet clear if these are merger shocks}. We therefore consider a weak merger scenario unlikely. Based on the available information on the dynamics of the cluster we cannot form a strong conclusion on the nature of this non-detection. Abell~3266 therefore joins a small collection of radio relic-hosting merging clusters without giant radio haloes.

\section{Summary}

Observations of the galaxy cluster Abell~3266 with the MWA have revealed three components of diffuse, non-thermal radio emission. Two fossil sources are reported for the first time, and a previously detected radio relic towards the southern periphery of the cluster is also observed. For the relic and the northern fossil source, patchy emission is seen at 887.5~MHz in the RACS images, though the fossil component associated with the complex WAT has no counterpart neither the RACS/SUMSS images nor the MGCLS images. The fossil sources are observed to have steep spectra across the MWA band (88--216~MHz), with $\alpha = -1.7 \pm 0.1$ and $\alpha = -2.4\pm 0.4$ for the northern and western fossils, respectively. The southern radio relic appears much flatter in spectrum with $\alpha = -1.04 \pm 0.09$, hinting towards shock-driven acceleration. Additionally, we model the low-frequency spectrum of the $\sim 630$~kpc radio galaxy north of the cluster centre, finding a steep spectrum with a spectral index of $\alpha=-1.26 \pm 0.22$, suggesting significant contribution from aged plasma components. 

No halo is detected above a peak surface brightness of \CORRS{$\sim 2$~\textmu Jy\,arcsec$^{-2}$} at 216~MHz, providing an upper limit to the 150-MHz luminosity of \CORRS{$P_{150~\text{MHz}}^{500~\text{kpc}} = 7.2 \times 10^{24}$~W\,Hz$^{-1}$ ($P_{150~\text{MHz}}^{1~\text{Mpc}} = 18 \times 10^{24}$~W\,Hz$^{-1}$)}, depending on the radius of the halo. These limits are \CORRS{up to $\sim 2$--5} times lower than expected from giant radio halo scaling relations at 150~MHz. Why a giant radio halo is undetected in Abell~3266 is unclear, as the cluster system appears in the exact dynamical state that would allow the formation of a giant radio halo. If a heretofore undetected radio halo is significantly under-luminous with respect to the $P_{\text{150~MHz}}$--$M$ scaling relations, further deep observations with MeerKAT -- or in the future with the upgraded MWA Phase III or SKA-Low -- may uncover it.

\section*{Acknowledgements}

The authors would like to thank the anonymous referee for their feedback so far that has improved the quality of this work. SWD acknowledges an Australian Government Research Training Program scholarship administered through Curtin University. CJR acknowledges financial support from the ERC Starting Grant `DRANOEL', number 714245.

Support for the operation of the MWA is provided by the Australian Government (NCRIS), under a contract to Curtin University administered by Astronomy Australia Limited. The Australian SKA Pathfinder is part of the Australia Telescope National Facility which is managed by CSIRO. Operation of ASKAP is funded by the Australian Government with support from the National Collaborative Research Infrastructure Strategy. ASKAP uses the resources of the Pawsey Supercomputing Centre. Establishment of ASKAP, the Murchison Radio-astronomy Observatory and the Pawsey Supercomputing Centre are initiatives of the Australian Government, with support from the Government of Western Australia and the Science and Industry Endowment Fund. We acknowledge the Wajarri Yamatji people as the traditional owners of the Observatory site. This paper includes archived data obtained through the CSIRO ASKAP Science Data Archive, CASDA (\url{https://data.csiro.au}). 

This research made use of a number of \textsc{python} packages not explicitly mentioned in the main text: \textsc{aplpy} \citep{Robitaille2012}, \textsc{astropy} \citep{Astropy2018}, \textsc{matplotlib} \citep{Hunter2007}, \textsc{numpy} \citep{Numpy2011}, \textsc{scipy} \citep{Jones2001}, and \textsc{cmasher} \citep{cmasher}.
\section*{Data Availability}

The the images underlying this article will be shared on reasonable request to the corresponding author, however, the raw data can be accessed through the MWA ASVO \footnote{All-Sky Virtual Observatory.} (\url{https://asvo.mwatelescope.org/}) for the MWA data. The ASKAP data is available through CASDA (\url{https://data.csiro.au}) as part of the RACS data collections (\url{http://hdl.handle.net/102.100.100/374842?index=1}) and the \xmm{} data is available through the Science Data Archive (\url{https://www.cosmos.esa.int/web/xmm-newton/xsa}).



\bibliographystyle{mnras}
\bibliography{references} 

\begin{thebibliography}{}
\makeatletter
\relax
\def\mn@urlcharsother{\let\do\@makeother \do\$\do\&\do\#\do\^\do\_\do\%\do\~}
\def\mn@doi{\begingroup\mn@urlcharsother \@ifnextchar [ {\mn@doi@}
  {\mn@doi@[]}}
\def\mn@doi@[#1]#2{\def\@tempa{#1}\ifx\@tempa\@empty \href
  {http://dx.doi.org/#2} {doi:#2}\else \href {http://dx.doi.org/#2} {#1}\fi
  \endgroup}
\def\mn@eprint#1#2{\mn@eprint@#1:#2::\@nil}
\def\mn@eprint@arXiv#1{\href {http://arxiv.org/abs/#1} {{\tt arXiv:#1}}}
\def\mn@eprint@dblp#1{\href {http://dblp.uni-trier.de/rec/bibtex/#1.xml}
  {dblp:#1}}
\def\mn@eprint@#1:#2:#3:#4\@nil{\def\@tempa {#1}\def\@tempb {#2}\def\@tempc
  {#3}\ifx \@tempc \@empty \let \@tempc \@tempb \let \@tempb \@tempa \fi \ifx
  \@tempb \@empty \def\@tempb {arXiv}\fi \@ifundefined
  {mn@eprint@\@tempb}{\@tempb:\@tempc}{\expandafter \expandafter \csname
  mn@eprint@\@tempb\endcsname \expandafter{\@tempc}}}

\bibitem[\protect\citeauthoryear{{Abell}, {Corwin}  \& {Olowin}}{{Abell}
  et~al.}{1989}]{aco89}
{Abell} G.~O.,  {Corwin} Jr. H.~G.,   {Olowin} R.~P.,  1989, \mn@doi [\apjs]
  {10.1086/191333}, \href {http://adsabs.harvard.edu/abs/1989ApJS...70....1A}
  {70, 1}

\bibitem[\protect\citeauthoryear{{Arras}, {Reinecke}, {Westermann}  \&
  {En{\ss}in}}{{Arras} et~al.}{2021}]{wgridder1}
{Arras} P.,  {Reinecke} M.,  {Westermann} R.,   {En{\ss}in} T.~A.,  2021,
  \mn@doi [\aap] {10.1051/0004-6361/202039723}, \href
  {https://ui.adsabs.harvard.edu/abs/2021A&A...646A..58A} {646, A58}

\bibitem[\protect\citeauthoryear{{Bagchi}, {Durret}, {Neto}  \&
  {Paul}}{{Bagchi} et~al.}{2006}]{Bagchi2006}
{Bagchi} J.,  {Durret} F.,  {Neto} G. B.~L.,   {Paul} S.,  2006, \mn@doi
  [Science] {10.1126/science.1131189}, \href
  {https://ui.adsabs.harvard.edu/abs/2006Sci...314..791B} {314, 791}

\bibitem[\protect\citeauthoryear{{Bartalucci} et~al.,}{{Bartalucci}
  et~al.}{2017}]{bartalucci2017}
{Bartalucci} I.,  et~al., 2017, \mn@doi [\aap] {10.1051/0004-6361/201731689},
  \href {https://ui.adsabs.harvard.edu/abs/2017A&A...608A..88B} {608, A88}

\bibitem[\protect\citeauthoryear{{Basu}}{{Basu}}{2012}]{basu12}
{Basu} K.,  2012, \mn@doi [\mnras] {10.1111/j.1745-3933.2012.01217.x}, \href
  {http://adsabs.harvard.edu/abs/2012MNRAS.421L.112B} {421, L112}

\bibitem[\protect\citeauthoryear{{Bernardi} et~al.,}{{Bernardi}
  et~al.}{2016}]{bvc+16}
{Bernardi} G.,  et~al., 2016, \mn@doi [\mnras] {10.1093/mnras/stv2589}, \href
  {http://adsabs.harvard.edu/abs/2016MNRAS.456.1259B} {456, 1259}

\bibitem[\protect\citeauthoryear{{Bock}, {Large}  \& {Sadler}}{{Bock}
  et~al.}{1999}]{bls99}
{Bock} D.~C.-J.,  {Large} M.~I.,   {Sadler} E.~M.,  1999, \mn@doi [\aj]
  {10.1086/300786}, \href {http://adsabs.harvard.edu/abs/1999AJ....117.1578B}
  {117, 1578}

\bibitem[\protect\citeauthoryear{{Bonafede}, {Giovannini}, {Feretti}, {Govoni}
  \& {Murgia}}{{Bonafede} et~al.}{2009}]{Bonafede2009}
{Bonafede} A.,  {Giovannini} G.,  {Feretti} L.,  {Govoni} F.,   {Murgia} M.,
  2009, \mn@doi [\aap] {10.1051/0004-6361:200810588}, \href
  {https://ui.adsabs.harvard.edu/abs/2009A&A...494..429B} {494, 429}

\bibitem[\protect\citeauthoryear{{Bonafede}, {Intema}, {Br{\"u}ggen},
  {Girardi}, {Nonino}, {Kantharia}, {van Weeren}  \&
  {R{\"o}ttgering}}{{Bonafede} et~al.}{2014}]{Bonafede2014}
{Bonafede} A.,  {Intema} H.~T.,  {Br{\"u}ggen} M.,  {Girardi} M.,  {Nonino} M.,
   {Kantharia} N.,  {van Weeren} R.~J.,   {R{\"o}ttgering} H.~J.~A.,  2014,
  \mn@doi [\apj] {10.1088/0004-637X/785/1/1}, \href
  {https://ui.adsabs.harvard.edu/abs/2014ApJ...785....1B} {785, 1}

\bibitem[\protect\citeauthoryear{{Bonafede} et~al.,}{{Bonafede}
  et~al.}{2017}]{Bonafede2017}
{Bonafede} A.,  et~al., 2017, \mn@doi [\mnras] {10.1093/mnras/stx1475}, \href
  {https://ui.adsabs.harvard.edu/abs/2017MNRAS.470.3465B} {470, 3465}

\bibitem[\protect\citeauthoryear{{Botteon} et~al.,}{{Botteon}
  et~al.}{2018}]{Botteon2018}
{Botteon} A.,  et~al., 2018, \mn@doi [\mnras] {10.1093/mnras/sty1102}, \href
  {https://ui.adsabs.harvard.edu/abs/2018MNRAS.478..885B} {478, 885}

\bibitem[\protect\citeauthoryear{{Botteon} et~al.,}{{Botteon}
  et~al.}{2021}]{Botteon2021}
{Botteon} A.,  et~al., 2021, \mn@doi [\aap] {10.1051/0004-6361/202040083},
  \href {https://ui.adsabs.harvard.edu/abs/2021A&A...649A..37B} {649, A37}

\bibitem[\protect\citeauthoryear{{Boxelaar}, {van Weeren}  \&
  {Botteon}}{{Boxelaar} et~al.}{2021}]{Boxelaar2021}
{Boxelaar} J.~M.,  {van Weeren} R.~J.,   {Botteon} A.,  2021, \mn@doi
  [Astronomy and Computing] {10.1016/j.ascom.2021.100464}, \href
  {https://ui.adsabs.harvard.edu/abs/2021A&C....3500464B} {35, 100464}

\bibitem[\protect\citeauthoryear{{Breuer}, {Werner}, {Mernier}, {Mroczkowski},
  {Simionescu}, {Clarke}, {ZuHone}  \& {Di Mascolo}}{{Breuer}
  et~al.}{2020}]{Breuer2020}
{Breuer} J.~P.,  {Werner} N.,  {Mernier} F.,  {Mroczkowski} T.,  {Simionescu}
  A.,  {Clarke} T.~E.,  {ZuHone} J.~A.,   {Di Mascolo} L.,  2020, \mn@doi
  [\mnras] {10.1093/mnras/staa1492}, \href
  {https://ui.adsabs.harvard.edu/abs/2020MNRAS.495.5014B} {495, 5014}

\bibitem[\protect\citeauthoryear{{Brunetti} \& {Jones}}{{Brunetti} \&
  {Jones}}{2014}]{bj14}
{Brunetti} G.,  {Jones} T.~W.,  2014, \mn@doi [International Journal of Modern
  Physics D] {10.1142/S0218271814300079}, \href
  {http://adsabs.harvard.edu/abs/2014IJMPD..2330007B} {23, 1430007}

\bibitem[\protect\citeauthoryear{{Brunetti} et~al.,}{{Brunetti}
  et~al.}{2008}]{bgc+08}
{Brunetti} G.,  et~al., 2008, \mn@doi [\nat] {10.1038/nature07379}, \href
  {http://adsabs.harvard.edu/abs/2008Natur.455..944B} {455, 944}

\bibitem[\protect\citeauthoryear{{Carretti} et~al.,}{{Carretti}
  et~al.}{2013}]{Carretti2013}
{Carretti} E.,  et~al., 2013, \mn@doi [\mnras] {10.1093/mnras/stt002}, \href
  {https://ui.adsabs.harvard.edu/abs/2013MNRAS.430.1414C} {430, 1414}

\bibitem[\protect\citeauthoryear{{Cassano}, {Brunetti}, {Setti}, {Govoni}  \&
  {Dolag}}{{Cassano} et~al.}{2007}]{Cassano2007}
{Cassano} R.,  {Brunetti} G.,  {Setti} G.,  {Govoni} F.,   {Dolag} K.,  2007,
  \mn@doi [\mnras] {10.1111/j.1365-2966.2007.11901.x}, \href
  {https://ui.adsabs.harvard.edu/abs/2007MNRAS.378.1565C} {378, 1565}

\bibitem[\protect\citeauthoryear{{Cassano}, {Ettori}, {Giacintucci},
  {Brunetti}, {Markevitch}, {Venturi}  \& {Gitti}}{{Cassano}
  et~al.}{2010}]{Cassano2010}
{Cassano} R.,  {Ettori} S.,  {Giacintucci} S.,  {Brunetti} G.,  {Markevitch}
  M.,  {Venturi} T.,   {Gitti} M.,  2010, \mn@doi [\apjl]
  {10.1088/2041-8205/721/2/L82}, \href
  {https://ui.adsabs.harvard.edu/abs/2010ApJ...721L..82C} {721, L82}

\bibitem[\protect\citeauthoryear{{Cassano} et~al.,}{{Cassano}
  et~al.}{2013}]{ceb+13}
{Cassano} R.,  et~al., 2013, \mn@doi [\apj] {10.1088/0004-637X/777/2/141},
  \href {http://adsabs.harvard.edu/abs/2013ApJ...777..141C} {777, 141}

\bibitem[\protect\citeauthoryear{{Cassano}, {Brunetti}, {Giocoli}  \&
  {Ettori}}{{Cassano} et~al.}{2016}]{Cassano2016}
{Cassano} R.,  {Brunetti} G.,  {Giocoli} C.,   {Ettori} S.,  2016, \mn@doi
  [\aap] {10.1051/0004-6361/201628414}, \href
  {https://ui.adsabs.harvard.edu/abs/2016A&A...593A..81C} {593, A81}

\bibitem[\protect\citeauthoryear{{Chapman}, {Dempsey}, {Miller}, {Heywood},
  {Pritchard}, {Sangster}, {Whiting}  \& {Dart}}{{Chapman}
  et~al.}{2017}]{casda}
{Chapman} J.~M.,  {Dempsey} J.,  {Miller} D.,  {Heywood} I.,  {Pritchard} J.,
  {Sangster} E.,  {Whiting} M.,   {Dart} M.,  2017, {CASDA: The CSIRO ASKAP
  Science Data Archive}.
p.~73

\bibitem[\protect\citeauthoryear{{Chokshi}, {Line}, {Barry}, {Ung}, {Kenney},
  {McPhail}, {Williams}  \& {Webster}}{{Chokshi} et~al.}{2021}]{Chokshi2021}
{Chokshi} A.,  {Line} J.~L.~B.,  {Barry} N.,  {Ung} D.,  {Kenney} D.,
  {McPhail} A.,  {Williams} A.,   {Webster} R.~L.,  2021, \mn@doi [\mnras]
  {10.1093/mnras/stab156}, \href
  {https://ui.adsabs.harvard.edu/abs/2021MNRAS.502.1990C} {502, 1990}

\bibitem[\protect\citeauthoryear{{Cuciti}, {Brunetti}, {van Weeren},
  {Bonafede}, {Dallacasa}, {Cassano}, {Venturi}  \& {Kale}}{{Cuciti}
  et~al.}{2018}]{Cuciti2018}
{Cuciti} V.,  {Brunetti} G.,  {van Weeren} R.,  {Bonafede} A.,  {Dallacasa} D.,
   {Cassano} R.,  {Venturi} T.,   {Kale} R.,  2018, \mn@doi [\aap]
  {10.1051/0004-6361/201731174}, \href
  {https://ui.adsabs.harvard.edu/abs/2018A&A...609A..61C} {609, A61}

\bibitem[\protect\citeauthoryear{{Cuciti} et~al.,}{{Cuciti}
  et~al.}{2021}]{Cuciti2021b}
{Cuciti} V.,  et~al., 2021, \mn@doi [\aap] {10.1051/0004-6361/202039208}, \href
  {https://ui.adsabs.harvard.edu/abs/2021A&A...647A..51C} {647, A51}

\bibitem[\protect\citeauthoryear{{Cypriano}, {Sodr{\'e}}, {Campusano}, {Kneib},
  {Giovanelli}, {Haynes}, {Dale}  \& {Hardy}}{{Cypriano}
  et~al.}{2001}]{cypriano01}
{Cypriano} E.~S.,  {Sodr{\'e}} Laerte J.,  {Campusano} L.~E.,  {Kneib} J.-P.,
  {Giovanelli} R.,  {Haynes} M.~P.,  {Dale} D.~A.,   {Hardy} E.,  2001, \mn@doi
  [\aj] {10.1086/318010}, \href
  {https://ui.adsabs.harvard.edu/abs/2001AJ....121...10C} {121, 10}

\bibitem[\protect\citeauthoryear{{Dehghan}}{{Dehghan}}{2014}]{dehghan-phd}
{Dehghan} S.,  2014, PhD thesis, Victoria University of Wellington, \url
  {https://tewaharoa.victoria.ac.nz/discovery/fulldisplay?docid=alma99179242120102386&context=L&vid=64VUW_INST:VUWNUI&search_scope=MyInst_and_CI&isFrbr=true&tab=all&lang=en}

\bibitem[\protect\citeauthoryear{{Dehghan}, {Johnston-Hollitt}, {Colless}  \&
  {Miller}}{{Dehghan} et~al.}{2017}]{Dehghan2017}
{Dehghan} S.,  {Johnston-Hollitt} M.,  {Colless} M.,   {Miller} R.,  2017,
  \mn@doi [\mnras] {10.1093/mnras/stx582}, \href
  {https://ui.adsabs.harvard.edu/abs/2017MNRAS.468.2645D} {468, 2645}

\bibitem[\protect\citeauthoryear{{Donnert}, {Dolag}, {Brunetti}  \&
  {Cassano}}{{Donnert} et~al.}{2013}]{ddbc13}
{Donnert} J.,  {Dolag} K.,  {Brunetti} G.,   {Cassano} R.,  2013, \mn@doi
  [\mnras] {10.1093/mnras/sts628}, \href
  {http://adsabs.harvard.edu/abs/2013MNRAS.429.3564D} {429, 3564}

\bibitem[\protect\citeauthoryear{{Duchesne}, {Johnston-Hollitt}, {Zhu}, {Wayth}
   \& {Line}}{{Duchesne} et~al.}{2020}]{Duchesne2020a}
{Duchesne} S.~W.,  {Johnston-Hollitt} M.,  {Zhu} Z.,  {Wayth} R.~B.,   {Line}
  J.~L.~B.,  2020, \mn@doi [\pasa] {10.1017/pasa.2020.29}, \href
  {https://ui.adsabs.harvard.edu/abs/2020PASA...37...37D} {37, e037}

\bibitem[\protect\citeauthoryear{{Duchesne}, {Johnston-Hollitt}, {Bartalucci},
  {Hodgson}  \& {Pratt}}{{Duchesne} et~al.}{2021a}]{Duchesne2020b}
{Duchesne} S.~W.,  {Johnston-Hollitt} M.,  {Bartalucci} I.,  {Hodgson} T.,
  {Pratt} G.~W.,  2021a, \mn@doi [\pasa] {10.1017/pasa.2020.51}, \href
  {https://ui.adsabs.harvard.edu/abs/2021PASA...38....5D} {38, e005}

\bibitem[\protect\citeauthoryear{{Duchesne}, {Johnston-Hollitt}, {Offringa},
  {Pratt}, {Zheng}  \& {Dehghan}}{{Duchesne} et~al.}{2021b}]{Duchesne2017}
{Duchesne} S.~W.,  {Johnston-Hollitt} M.,  {Offringa} A.~R.,  {Pratt} G.~W.,
  {Zheng} Q.,   {Dehghan} S.,  2021b, \mn@doi [\pasa] {10.1017/pasa.2021.7},
  \href {https://ui.adsabs.harvard.edu/abs/2021PASA...38...10D} {38, e010}

\bibitem[\protect\citeauthoryear{{Duchesne}, {Johnston-Hollitt}  \&
  {Wilber}}{{Duchesne} et~al.}{2021c}]{Duchesne2021a}
{Duchesne} S.~W.,  {Johnston-Hollitt} M.,   {Wilber} A.~G.,  2021c, \mn@doi
  [\pasa] {10.1017/pasa.2021.24}, \href
  {https://ui.adsabs.harvard.edu/abs/2021PASA...38...31D} {38, e031}

\bibitem[\protect\citeauthoryear{{Duchesne}, {Johnston-Hollitt}  \&
  {Bartalucci}}{{Duchesne} et~al.}{2021d}]{Duchesne2021b}
{Duchesne} S.~W.,  {Johnston-Hollitt} M.,   {Bartalucci} I.,  2021d, \mn@doi
  [\pasa] {10.1017/pasa.2021.45}, \href
  {https://ui.adsabs.harvard.edu/abs/2021arXiv210612281D} {38, e053}

\bibitem[\protect\citeauthoryear{{Ettori} et~al.,}{{Ettori}
  et~al.}{2019}]{Ettori2019}
{Ettori} S.,  et~al., 2019, \mn@doi [\aap] {10.1051/0004-6361/201833323}, \href
  {https://ui.adsabs.harvard.edu/abs/2019A&A...621A..39E} {621, A39}

\bibitem[\protect\citeauthoryear{{Feretti}}{{Feretti}}{2000}]{Feretti2000}
{Feretti} L.,  2000, {Observational Properties of Diffuse Halos in Clusters}
  (\mn@eprint {arXiv} {astro-ph/0006379})

\bibitem[\protect\citeauthoryear{{Feretti}, {Fusco-Femiano}, {Giovannini}  \&
  {Govoni}}{{Feretti} et~al.}{2001}]{ffgg01}
{Feretti} L.,  {Fusco-Femiano} R.,  {Giovannini} G.,   {Govoni} F.,  2001,
  \mn@doi [\aap] {10.1051/0004-6361:20010581}, \href
  {http://adsabs.harvard.edu/abs/2001A%26A...373..106F} {373, 106}

\bibitem[\protect\citeauthoryear{{Finoguenov}, {Henriksen}, {Miniati}, {Briel}
  \& {Jones}}{{Finoguenov} et~al.}{2006}]{Finoguenov2006}
{Finoguenov} A.,  {Henriksen} M.~J.,  {Miniati} F.,  {Briel} U.~G.,   {Jones}
  C.,  2006, \mn@doi [\apj] {10.1086/503285}, \href
  {https://ui.adsabs.harvard.edu/abs/2006ApJ...643..790F} {643, 790}

\bibitem[\protect\citeauthoryear{{Fleenor}, {Rose}, {Christiansen}, {Hunstead},
  {Johnston-Hollitt}, {Drinkwater}  \& {Saunders}}{{Fleenor}
  et~al.}{2005}]{fleenor05}
{Fleenor} M.~C.,  {Rose} J.~A.,  {Christiansen} W.~A.,  {Hunstead} R.~W.,
  {Johnston-Hollitt} M.,  {Drinkwater} M.~J.,   {Saunders} W.,  2005, \mn@doi
  [\aj] {10.1086/431972}, \href
  {https://ui.adsabs.harvard.edu/abs/2005AJ....130..957F} {130, 957}

\bibitem[\protect\citeauthoryear{{Giacintucci}, {Markevitch}, {Venturi},
  {Clarke}, {Cassano}  \& {Mazzotta}}{{Giacintucci}
  et~al.}{2014}]{Giacintucci2014a}
{Giacintucci} S.,  {Markevitch} M.,  {Venturi} T.,  {Clarke} T.~E.,  {Cassano}
  R.,   {Mazzotta} P.,  2014, \mn@doi [\apj] {10.1088/0004-637X/781/1/9}, \href
  {https://ui.adsabs.harvard.edu/abs/2014ApJ...781....9G} {781, 9}

\bibitem[\protect\citeauthoryear{{Gitti}, {Brunetti}  \& {Setti}}{{Gitti}
  et~al.}{2002}]{Gitti2002}
{Gitti} M.,  {Brunetti} G.,   {Setti} G.,  2002, \mn@doi [\aap]
  {10.1051/0004-6361:20020284}, \href
  {https://ui.adsabs.harvard.edu/abs/2002A&A...386..456G} {386, 456}

\bibitem[\protect\citeauthoryear{{Govoni} \& {Feretti}}{{Govoni} \&
  {Feretti}}{2004}]{gf04}
{Govoni} F.,  {Feretti} L.,  2004, \mn@doi [International Journal of Modern
  Physics D] {10.1142/S0218271804005080}, \href
  {http://adsabs.harvard.edu/abs/2004IJMPD..13.1549G} {13, 1549}

\bibitem[\protect\citeauthoryear{{Hindson} et~al.,}{{Hindson}
  et~al.}{2014}]{Hindson2014}
{Hindson} L.,  et~al., 2014, \mn@doi [\mnras] {10.1093/mnras/stu1669}, \href
  {https://ui.adsabs.harvard.edu/abs/2014MNRAS.445..330H} {445, 330}

\bibitem[\protect\citeauthoryear{{Hodgson}, {Vazza}, {Johnston-Hollitt},
  {Duchesne}  \& {McKinley}}{{Hodgson} et~al.}{2021a}]{Hodgson2021b}
{Hodgson} T.,  {Vazza} F.,  {Johnston-Hollitt} M.,  {Duchesne} S.~W.,
  {McKinley} B.,  2021a, {Stacking the Synchrotron Cosmic Web with FIGARO}
  (\mn@eprint {arXiv} {2108.13682})

\bibitem[\protect\citeauthoryear{{Hodgson}, {Bartalucci}, {Johnston-Hollitt},
  {McKinley}, {Vazza}  \& {Wittor}}{{Hodgson} et~al.}{2021b}]{Hodgson2021}
{Hodgson} T.,  {Bartalucci} I.,  {Johnston-Hollitt} M.,  {McKinley} B.,
  {Vazza} F.,   {Wittor} D.,  2021b, \mn@doi [\apj] {10.3847/1538-4357/abe384},
  \href {https://ui.adsabs.harvard.edu/abs/2021ApJ...909..198H} {909, 198}

\bibitem[\protect\citeauthoryear{{Hotan}, {Whiting}, {Huynh}  \&
  {Moss}}{{Hotan} et~al.}{2020}]{askap:racs}
{Hotan} A.,  {Whiting} M.,  {Huynh} M.,   {Moss} V.,  2020, ASKAP Data Products
  for Project AS113 (Other ASKAP pilot science including tests, TOOs or guest
  observations): images and visibilities. v1. CSIRO. Data Collection., \url
  {http://hdl.handle.net/102.100.100/348894?index=1}

\bibitem[\protect\citeauthoryear{{Hotan} et~al.,}{{Hotan}
  et~al.}{2021}]{Hotan2021}
{Hotan} A.~W.,  et~al., 2021, \mn@doi [\pasa] {10.1017/pasa.2021.1}, \href
  {https://ui.adsabs.harvard.edu/abs/2021PASA...38....9H} {38, e009}

\bibitem[\protect\citeauthoryear{{Hunter}}{{Hunter}}{2007}]{Hunter2007}
{Hunter} J.~D.,  2007, \mn@doi [Computing in Science and Engineering]
  {10.1109/MCSE.2007.55}, \href
  {http://adsabs.harvard.edu/abs/2007CSE.....9...90H} {9, 90}

\bibitem[\protect\citeauthoryear{{Hurley-Walker} et~al.,}{{Hurley-Walker}
  et~al.}{2017}]{gleamegc}
{Hurley-Walker} N.,  et~al., 2017, \mn@doi [\mnras] {10.1093/mnras/stw2337},
  \href {http://adsabs.harvard.edu/abs/2017MNRAS.464.1146H} {464, 1146}

\bibitem[\protect\citeauthoryear{{Huynh}, {Dempsey}, {Whiting}  \&
  {Ophel}}{{Huynh} et~al.}{2020}]{Huynh2020}
{Huynh} M.,  {Dempsey} J.,  {Whiting} M.~T.,   {Ophel} M.,  2020, in
  {Ballester} P.,  {Ibsen} J.,  {Solar} M.,   {Shortridge} K.,  eds,
  Astronomical Society of the Pacific Conference Series Vol. 522, Astronomical
  Data Analysis Software and Systems XXVII. p.~263

\bibitem[\protect\citeauthoryear{{Johnston-Hollitt}}{{Johnston-Hollitt}}{2003}]{mj-h}
{Johnston-Hollitt} M.,  2003, PhD thesis, University of Adelaide, \url
  {http://hdl.handle.net/2440/21954}

\bibitem[\protect\citeauthoryear{Jones, Oliphant, Peterson  et~al.}{Jones
  et~al.}{2001}]{Jones2001}
Jones E.,  Oliphant T.,  Peterson P.,   et~al., 2001, {SciPy}: Open source
  scientific tools for {Python}, \url {http://www.scipy.org/}

\bibitem[\protect\citeauthoryear{{Jones} et~al.,}{{Jones}
  et~al.}{2021}]{Jones2021}
{Jones} A.,  et~al., 2021, \mn@doi [\mnras] {10.1093/mnras/stab1443}, \href
  {https://ui.adsabs.harvard.edu/abs/2021MNRAS.505.4762J} {505, 4762}

\bibitem[\protect\citeauthoryear{{Knowles} et~al.,}{{Knowles}
  et~al.}{2022}]{mgcls}
{Knowles} K.,  et~al., 2022, \mn@doi [\aap] {10.1051/0004-6361/202141488},
  \href {https://ui.adsabs.harvard.edu/abs/2022A&A...657A..56K} {657, A56}

\bibitem[\protect\citeauthoryear{{Liang}, {Hunstead}, {Birkinshaw}  \&
  {Andreani}}{{Liang} et~al.}{2000}]{lhba00}
{Liang} H.,  {Hunstead} R.~W.,  {Birkinshaw} M.,   {Andreani} P.,  2000,
  \mn@doi [\apj] {10.1086/317223}, \href
  {http://adsabs.harvard.edu/abs/2000ApJ...544..686L} {544, 686}

\bibitem[\protect\citeauthoryear{{Mandal} et~al.,}{{Mandal}
  et~al.}{2020}]{Mandal2020}
{Mandal} S.,  et~al., 2020, \mn@doi [\aap] {10.1051/0004-6361/201936560}, \href
  {https://ui.adsabs.harvard.edu/abs/2020A&A...634A...4M} {634, A4}

\bibitem[\protect\citeauthoryear{{Martinez Aviles} et~al.,}{{Martinez Aviles}
  et~al.}{2016}]{mfj+16}
{Martinez Aviles} G.,  et~al., 2016, \mn@doi [\aap]
  {10.1051/0004-6361/201628788}, \href
  {http://adsabs.harvard.edu/abs/2016A%26A...595A.116M} {595, A116}

\bibitem[\protect\citeauthoryear{{McConnell} et~al.,}{{McConnell}
  et~al.}{2020}]{racs1}
{McConnell} D.,  et~al., 2020, \mn@doi [\pasa] {10.1017/pasa.2020.41}, \href
  {https://ui.adsabs.harvard.edu/abs/2020PASA...37...48M} {37, e048}

\bibitem[\protect\citeauthoryear{{Miller}}{{Miller}}{2012}]{Miller2012}
{Miller} R.,  2012, Bachelor's thesis, Victoria University of Wellington

\bibitem[\protect\citeauthoryear{{Murgia}, {Govoni}, {Markevitch}, {Feretti},
  {Giovannini}, {Taylor}  \& {Carretti}}{{Murgia} et~al.}{2009}]{Murgia2009}
{Murgia} M.,  {Govoni} F.,  {Markevitch} M.,  {Feretti} L.,  {Giovannini} G.,
  {Taylor} G.~B.,   {Carretti} E.,  2009, \mn@doi [\aap]
  {10.1051/0004-6361/200911659}, \href
  {https://ui.adsabs.harvard.edu/abs/2009A&A...499..679M} {499, 679}

\bibitem[\protect\citeauthoryear{{Murgia}, {Govoni}, {Feretti}  \&
  {Giovannini}}{{Murgia} et~al.}{2010}]{Murgia2010}
{Murgia} M.,  {Govoni} F.,  {Feretti} L.,   {Giovannini} G.,  2010, \mn@doi
  [\aap] {10.1051/0004-6361/200913414}, \href
  {https://ui.adsabs.harvard.edu/abs/2010A&A...509A..86M} {509, A86}

\bibitem[\protect\citeauthoryear{{Murgia} et~al.,}{{Murgia}
  et~al.}{2011}]{mpm+11}
{Murgia} M.,  et~al., 2011, \mn@doi [\aap] {10.1051/0004-6361/201015302}, \href
  {http://adsabs.harvard.edu/abs/2011A%26A...526A.148M} {526, A148}

\bibitem[\protect\citeauthoryear{{Murphy}}{{Murphy}}{1999}]{murphy99}
{Murphy} T.,  1999, PhD thesis, University of Sydney, \url
  {http://www.astrop.physics.usyd.edu.au/RELICS/thesis/thesis.html}

\bibitem[\protect\citeauthoryear{{Offringa} \& {Smirnov}}{{Offringa} \&
  {Smirnov}}{2017}]{wsclean2}
{Offringa} A.~R.,  {Smirnov} O.,  2017, \mn@doi [\mnras]
  {10.1093/mnras/stx1547}, \href
  {https://ui.adsabs.harvard.edu/abs/2017MNRAS.471..301O} {471, 301}

\bibitem[\protect\citeauthoryear{{Offringa} et~al.,}{{Offringa}
  et~al.}{2014}]{wsclean1}
{Offringa} A.~R.,  et~al., 2014, \mn@doi [\mnras] {10.1093/mnras/stu1368},
  \href {https://ui.adsabs.harvard.edu/abs/2014MNRAS.444..606O} {444, 606}

\bibitem[\protect\citeauthoryear{{Ogrean} et~al.,}{{Ogrean}
  et~al.}{2015}]{Ogrean2015}
{Ogrean} G.~A.,  et~al., 2015, \mn@doi [\apj] {10.1088/0004-637X/812/2/153},
  \href {https://ui.adsabs.harvard.edu/abs/2015ApJ...812..153O} {812, 153}

\bibitem[\protect\citeauthoryear{{Orr{\'u}}, {Murgia}, {Feretti}, {Govoni},
  {Brunetti}, {Giovannini}, {Girardi}  \& {Setti}}{{Orr{\'u}}
  et~al.}{2007}]{Orru2007}
{Orr{\'u}} E.,  {Murgia} M.,  {Feretti} L.,  {Govoni} F.,  {Brunetti} G.,
  {Giovannini} G.,  {Girardi} M.,   {Setti} G.,  2007, \mn@doi [\aap]
  {10.1051/0004-6361:20066118}, \href
  {https://ui.adsabs.harvard.edu/abs/2007A&A...467..943O} {467, 943}

\bibitem[\protect\citeauthoryear{{Pearce} et~al.,}{{Pearce}
  et~al.}{2017}]{Pearce2017}
{Pearce} C.~J.~J.,  et~al., 2017, \mn@doi [\apj] {10.3847/1538-4357/aa7e2f},
  \href {https://ui.adsabs.harvard.edu/abs/2017ApJ...845...81P} {845, 81}

\bibitem[\protect\citeauthoryear{{Piffaretti}, {Arnaud}, {Pratt},
  {Pointecouteau}  \& {Melin}}{{Piffaretti} et~al.}{2011}]{pap+11}
{Piffaretti} R.,  {Arnaud} M.,  {Pratt} G.~W.,  {Pointecouteau} E.,   {Melin}
  J.-B.,  2011, \mn@doi [\aap] {10.1051/0004-6361/201015377}, \href
  {http://adsabs.harvard.edu/abs/2011A%26A...534A.109P} {534, A109}

\bibitem[\protect\citeauthoryear{{Planck Collaboration} et~al.,}{{Planck
  Collaboration} et~al.}{2016}]{planck16}
{Planck Collaboration} et~al., 2016, \mn@doi [\aap]
  {10.1051/0004-6361/201525823}, \href
  {https://ui.adsabs.harvard.edu/abs/2016A&A...594A..27P} {594, A27}

\bibitem[\protect\citeauthoryear{{Predehl} et~al.,}{{Predehl}
  et~al.}{2021}]{erosita2}
{Predehl} P.,  et~al., 2021, \mn@doi [\aap] {10.1051/0004-6361/202039313},
  \href {https://ui.adsabs.harvard.edu/abs/2021A&A...647A...1P} {647, A1}

\bibitem[\protect\citeauthoryear{{Quici} et~al.,}{{Quici}
  et~al.}{2021}]{Quici2021}
{Quici} B.,  et~al., 2021, \mn@doi [\pasa] {10.1017/pasa.2020.49}, \href
  {https://ui.adsabs.harvard.edu/abs/2021PASA...38....8Q} {38, e008}

\bibitem[\protect\citeauthoryear{{Quintana}, {Ramirez}  \& {Way}}{{Quintana}
  et~al.}{1996}]{Quintana1996}
{Quintana} H.,  {Ramirez} A.,   {Way} M.~J.,  1996, \mn@doi [\aj]
  {10.1086/117987}, \href
  {https://ui.adsabs.harvard.edu/abs/1996AJ....112...36Q} {112, 36}

\bibitem[\protect\citeauthoryear{{Reichardt} et~al.,}{{Reichardt}
  et~al.}{2013}]{spt2}
{Reichardt} C.~L.,  et~al., 2013, \mn@doi [\apj] {10.1088/0004-637X/763/2/127},
  \href {http://cdsads.u-strasbg.fr/abs/2013ApJ...763..127R} {763, 127}

\bibitem[\protect\citeauthoryear{Riseley}{Riseley}{2016}]{riseley-phd}
Riseley C.~J.,  2016, PhD thesis, University of Southampton, \url
  {https://eprints.soton.ac.uk/405434/}

\bibitem[\protect\citeauthoryear{{Riseley}, {Scaife}, {Oozeer}, {Magnus}  \&
  {Wise}}{{Riseley} et~al.}{2015}]{Riseley2015}
{Riseley} C.~J.,  {Scaife} A.~M.~M.,  {Oozeer} N.,  {Magnus} L.,   {Wise}
  M.~W.,  2015, \mn@doi [\mnras] {10.1093/mnras/stu2591}, \href
  {http://adsabs.harvard.edu/abs/2015MNRAS.447.1895R} {447, 1895}

\bibitem[\protect\citeauthoryear{{Robertson} \& {Roach}}{{Robertson} \&
  {Roach}}{1990}]{Robertson1990}
{Robertson} J.~G.,  {Roach} G.~J.,  1990, \mnras, \href
  {https://ui.adsabs.harvard.edu/abs/1990MNRAS.247..387R} {247, 387}

\bibitem[\protect\citeauthoryear{{Robitaille} \& {Bressert}}{{Robitaille} \&
  {Bressert}}{2012}]{Robitaille2012}
{Robitaille} T.,  {Bressert} E.,  2012, {APLpy: Astronomical Plotting Library
  in Python}, Astrophysics Source Code Library (\mn@eprint {ascl} {1208.017})

\bibitem[\protect\citeauthoryear{{Rudnick}, {Cotton}, {Knowles}  \&
  {Kolokythas}}{{Rudnick} et~al.}{2021}]{Rudnick2021}
{Rudnick} L.,  {Cotton} W.,  {Knowles} K.,   {Kolokythas} K.,  2021, \mn@doi
  [Galaxies] {10.3390/galaxies9040081}, \href
  {https://ui.adsabs.harvard.edu/abs/2021Galax...9...81R} {9, 81}

\bibitem[\protect\citeauthoryear{{Sanders} et~al.,}{{Sanders}
  et~al.}{2021}]{Sanders2021}
{Sanders} J.~S.,  et~al., 2021, {Studying the merging cluster Abell 3266 with
  eROSITA} (\mn@eprint {arXiv} {2106.14534})

\bibitem[\protect\citeauthoryear{{Sauvageot}, {Belsole}  \&
  {Pratt}}{{Sauvageot} et~al.}{2005}]{Sauvageot2005}
{Sauvageot} J.~L.,  {Belsole} E.,   {Pratt} G.~W.,  2005, \mn@doi [\aap]
  {10.1051/0004-6361:20053242}, \href
  {https://ui.adsabs.harvard.edu/abs/2005A&A...444..673S} {444, 673}

\bibitem[\protect\citeauthoryear{{Shakouri}, {Johnston-Hollitt}  \&
  {Pratt}}{{Shakouri} et~al.}{2016}]{sjp16}
{Shakouri} S.,  {Johnston-Hollitt} M.,   {Pratt} G.~W.,  2016, \mn@doi [\mnras]
  {10.1093/mnras/stw812}, \href
  {http://adsabs.harvard.edu/abs/2016MNRAS.459.2525S} {459, 2525}

\bibitem[\protect\citeauthoryear{{Slee}, {Roy}, {Murgia}, {Andernach}  \&
  {Ehle}}{{Slee} et~al.}{2001}]{srm+01}
{Slee} O.~B.,  {Roy} A.~L.,  {Murgia} M.,  {Andernach} H.,   {Ehle} M.,  2001,
  \mn@doi [\aj] {10.1086/322105}, \href
  {http://adsabs.harvard.edu/abs/2001AJ....122.1172S} {122, 1172}

\bibitem[\protect\citeauthoryear{{The Astropy Collaboration} et~al.,}{{The
  Astropy Collaboration} et~al.}{2018}]{Astropy2018}
{The Astropy Collaboration} et~al., 2018, \mn@doi [\aj]
  {10.3847/1538-3881/aabc4f}, \href
  {https://ui.adsabs.harvard.edu/abs/2018AJ....156..123T} {156, 123}

\bibitem[\protect\citeauthoryear{{Tingay} et~al.,}{{Tingay}
  et~al.}{2013}]{tgb+13}
{Tingay} S.~J.,  et~al., 2013, \mn@doi [\pasa] {10.1017/pasa.2012.007}, \href
  {http://adsabs.harvard.edu/abs/2013PASA...30....7T} {30, 7}

\bibitem[\protect\citeauthoryear{{Wayth} et~al.,}{{Wayth}
  et~al.}{2015}]{wlb+15}
{Wayth} R.~B.,  et~al., 2015, \mn@doi [\pasa] {10.1017/pasa.2015.26}, \href
  {http://adsabs.harvard.edu/abs/2015PASA...32...25W} {32, 25}

\bibitem[\protect\citeauthoryear{{Wayth} et~al.,}{{Wayth}
  et~al.}{2018}]{wtt+18}
{Wayth} R.~B.,  et~al., 2018, \mn@doi [\pasa] {10.1017/pasa.2018.37}, \href
  {http://adsabs.harvard.edu/abs/2018PASA...35...33W} {35}

\bibitem[\protect\citeauthoryear{{Wilber} et~al.,}{{Wilber}
  et~al.}{2018}]{Wilber2018}
{Wilber} A.,  et~al., 2018, \mn@doi [\mnras] {10.1093/mnras/stx2568}, \href
  {https://ui.adsabs.harvard.edu/abs/2018MNRAS.473.3536W} {473, 3536}

\bibitem[\protect\citeauthoryear{{Wilber}, {Johnston-Hollitt}, {Duchesne},
  {Tasse}, {Akamatsu}, {Intema}  \& {Hodgson}}{{Wilber}
  et~al.}{2020}]{Wilber2020}
{Wilber} A.~G.,  {Johnston-Hollitt} M.,  {Duchesne} S.~W.,  {Tasse} C.,
  {Akamatsu} H.,  {Intema} H.,   {Hodgson} T.,  2020, \mn@doi [\pasa]
  {10.1017/pasa.2020.34}, \href
  {https://ui.adsabs.harvard.edu/abs/2020PASA...37...40W} {37, e040}

\bibitem[\protect\citeauthoryear{{Ye}, {Gull}, {Tan}  \& {Nikolic}}{{Ye}
  et~al.}{2021}]{wgridder2}
{Ye} H.,  {Gull} S.~F.,  {Tan} S.~M.,   {Nikolic} B.,  2021, \mn@doi [\mnras]
  {10.1093/mnras/stab3548}, \href
  {https://ui.adsabs.harvard.edu/abs/2021MNRAS.tmp.3220Y} {}

\bibitem[\protect\citeauthoryear{de Gasperin et~al.,}{de~Gasperin
  et~al.}{2021}]{deGasperin2021b}
de Gasperin F.,  et~al., 2021, MeerKAT view of the diffuse radio sources in
  Abell 3667 and their interactions with the thermal plasma (\mn@eprint {arXiv}
  {2111.06940})

\bibitem[\protect\citeauthoryear{{van Weeren} et~al.,}{{van Weeren}
  et~al.}{2016}]{vanWeeren2016}
{van Weeren} R.~J.,  et~al., 2016, \mn@doi [\apj]
  {10.3847/0004-637X/818/2/204}, \href
  {https://ui.adsabs.harvard.edu/abs/2016ApJ...818..204V} {818, 204}

\bibitem[\protect\citeauthoryear{{van Weeren} et~al.,}{{van Weeren}
  et~al.}{2017}]{vanWeeren2017}
{van Weeren} R.~J.,  et~al., 2017, \mn@doi [Nature Astronomy]
  {10.1038/s41550-016-0005}, \href
  {https://ui.adsabs.harvard.edu/abs/2017NatAs...1E...5V} {1, 0005}

\bibitem[\protect\citeauthoryear{{van Weeren}, {de Gasperin}, {Akamatsu},
  {Br{\"u}ggen}, {Feretti}, {Kang}, {Stroe}  \& {Zandanel}}{{van Weeren}
  et~al.}{2019}]{vda+19}
{van Weeren} R.~J.,  {de Gasperin} F.,  {Akamatsu} H.,  {Br{\"u}ggen} M.,
  {Feretti} L.,  {Kang} H.,  {Stroe} A.,   {Zandanel} F.,  2019, \mn@doi [\ssr]
  {10.1007/s11214-019-0584-z}, \href
  {http://adsabs.harvard.edu/abs/2019SSRv..215...16V} {215, 16}

\bibitem[\protect\citeauthoryear{{van Weeren} et~al.,}{{van Weeren}
  et~al.}{2021}]{vanWeeren2020}
{van Weeren} R.~J.,  et~al., 2021, \mn@doi [\aap]
  {10.1051/0004-6361/202039826}, \href
  {https://ui.adsabs.harvard.edu/abs/2021A&A...651A.115V} {651, A115}

\bibitem[\protect\citeauthoryear{van~der Velden}{van~der
  Velden}{2020}]{cmasher}
van~der Velden E.,  2020, \mn@doi [Journal of Open Source Software]
  {10.21105/joss.02004}, 5, 2004

\bibitem[\protect\citeauthoryear{van~der Walt, Colbert  \& Varoquaux}{van~der
  Walt et~al.}{2011}]{Numpy2011}
van~der Walt S.,  Colbert S.~C.,   Varoquaux G.,  2011, \mn@doi [Computing in
  Science Engineering] {10.1109/MCSE.2011.37}, 13, 22

\makeatother
\end{thebibliography}






\bsp	
\label{lastpage}
\end{document}